\documentclass[preprint2]{emulateapj}

\newcommand{\mr}{\mathrm}

\shorttitle{RHESSI Microflare Statistics. II.}
\shortauthors{Hannah et al.}
\begin{document}
\title{{\it RHESSI} Microflare Statistics II. \\
X-ray Imaging, Spectroscopy \& Energy Distributions.}

\author{I. G. Hannah, S. Christe\altaffilmark{1}, S. Krucker,
G. J. Hurford, H. S. Hudson, R. P. Lin\altaffilmark{1}}

\affil{Space Sciences Laboratory, University of California at Berkeley,
\\Berkeley, CA, 94720-7450, USA}

\email{hannah@ssl.berkeley.edu, schriste@ssl.berkeley.edu,\\
krucker@ssl.berkeley.edu, ghurford@ssl.berkeley.edu,\\
hhudson@ssl.berkeley.edu, rlin@ssl.berkeley.edu}

 \altaffiltext{1}{Physics Department,
 University of California at Berkeley,
 Berkeley, CA, 94720-7450, USA}

\begin{abstract} We present X-ray imaging and spectral analysis of all
microflares the Reuven Ramaty High Energy Solar Spectroscopic Imager
({\it RHESSI}) observed between March 2002 and March 2007, a total of
25,705 events. These microflares are small flares, from low {\it GOES} C
Class to below A Class (background subtracted) and are associated with
active regions. They were found by searching the 6-12 keV energy range
during periods when the full sensitivity of {\it RHESSI's} detectors was
available (see paper I). Each microflare is automatically analyzed at the
peak time of the 6-12~keV emission: the thermal source size is found by
forward-fitting the complex visibilities for 4-8 keV, and the spectral
parameters (temperature, emission measure, power-law index) are found
by forward fitting a thermal plus non-thermal model. The combination of
these parameters allows us to present the first statistical analysis of the
thermal and non-thermal energy at the peak times of microflares. On
average a {\it RHESSI} microflare has a fitted thermal loop width 8 Mm
(11$''$), length 23 Mm (32$''$) and volume $1\times10^{27}$~cm$^3$,
temperature 13~MK, emission measure $3\times 10^{46}$~cm$^{-3}$ and
density of $6\times10^9$~cm$^{-3}$. There is no correlation between the
loop size and the flare magnitude, either flux in the loop or GOES class,
indicating that microflares are not necessarily spatially small. There is also
no clear correlation between the thermal parameters except between the
{\it RHESSI} and {\it GOES} emission measures, the {\it GOES} values are
generally twice the {\it RHESSI} emission measures. The microflare thermal
energy at the time of peak emission in 6-12~keV ranges over $10^{26}$ to
$10^{30}$~erg and has a median value of $10^{28}$~erg. The frequency
distribution of the thermal energy deviates from a power-law at low and
high energies arising from a deficiency of events due to instrumental and
selection effects. It is difficult to compare this energy distribution to
previous thermal energy distributions of transient events, as the work
sought nanoflares through imaging in EUV or soft X-rays and covered just a
few hours. There are large uncertainties in the majority of the non-thermal
parameters, due to the steep spectra down to low energies. We typically
find a power-law index of $7$ above a break energy of $9$~keV, which
corresponds to a low-energy cut-off in the electron distribution as low as
$12$~keV. The resulting non-thermal power estimates, covering $10^{25}$
to $10^{28}$~erg~s$^{-1}$ with median value of $10^{26}$~erg~s$^{-1}$,
therefore have large uncertainties as well. The few microflares with
unexpectedly large non-thermal powers $10^{28}$~erg~s$^{-1}$ have the
smallest uncertainties, of about $10\%$. The total non-thermal energy
however is still small compared to that of large flares as it occurs for
shorter durations.\end{abstract}


\section{Introduction} The solar corona exhibits a myriad of transient
energy releases over many scales, from large flares down to nanoflares.
The frequency distribution of the energy in these events has been studied
extensively \citep{crosby1993,shimizu1995,krucker1998,asch2000,
parnell2000,lin2001,asch_parnell2002,bk2002} and has been found to be
well represented by a power-law of the form
 \begin{equation}\label{fig:pl}
 \mr{d}N=AW^{-\alpha}\mr{d}W
 \end{equation}
\noindent where $\mr{d}N$ is the number of events per unit time with
energy between $W$ and $W+\mr{d}W$. These distributions are of
particular interest as they elucidate the amount of energy available, how
often it is released and in which events and form (thermal or non-thermal)
it predominantly occurs. The energy release observed in normal-size flares
is not sufficient to constantly and consistently heat the corona to the
observed few million Kelvin. So the question is then whether this could be
achieved by extending the observed flare-like energy releases to smaller
scales. This concept can be expressed in terms of the power-law index of
this distribution: if $\alpha \geq 2$ then the smallest events have a high
occurrence rate and their energy dominates over larger flares, possibly
matching the energy in coronal heating \citep{hudson1991}. This requires
the assumption that these distributions be continuous into the
unobservable low-energy range, which is difficult to determine as there
are instrumental and selection effects that both cut-off and bias the
observed distribution \citep{asch_parnell2002}.

The instantaneous thermal energy in these events may be calculated from
\begin{equation}\label{eq:therm1} W_\mr{T}=3n_\mr{e}k_\mr{B}T V
\end{equation} \noindent where $n_\mr{e}$ is the electron density, $V$
the volume of the emitting thermal plasma, $k_\mr{B}$ Boltzmann's
constant, and $T$ the temperature. An estimate of the volume can be
obtained by imaging the events, however there can be an overestimate in
this observed volume to the true volume by a filling factor $f\approx1$ to
$10^{-4}$ \citep{cargill1997,takahashi2000}. In this work we assume
$f=1$, which will be discussed later. The temperature and emission
measure may be found either directly from the spectrum or by imaging
with different wavelength filters. Assuming constant density, the emission
measure is related to the density and volume as $EM=n_\mr{e}^2V$ and
so the thermal energy is \begin{equation}\label{eq:therm}
W_\mr{T}=3\sqrt{EM\cdot\,V}k_\mr{B}T. \end{equation} \noindent Note
that as losses are not taken into account here, the energy going into the
thermal plasma will be larger. For the smallest events this thermal energy
has been found using pixelated detectors in EUV and soft X-rays, with
simultaneous pixel brightenings within some area being registered as an
event
\citep{shimizu1995,krucker1998,asch2000,parnell2000,asch_parnell2002,bk2002}.
The area inferred from these, often spatially discontinuous, brightened
pixels gives an estimate of the volume, and observations using different
filters give temperature and emission measure information.

The events observed in EUV are termed ``nano''-flares as they have about
$10^{-9}$ times the energy in large flares, the limit of their observability
being down to $10^{24}$~erg \citep{asch2000}. Parker's hypothetical
nanoflare \citep{parker1988} is an estimate of the basic unit of a localized
impulsive burst of energy release, with energies $<10^{24}$~erg, with
ensembles of them constituting the observed events. The thermal energy
of these events outside of active regions has been investigated in soft
X-rays with {\it Yohkoh/SXT} \citep{krucker1997} finding energies of
$10^{25}$~erg per event, and in EUV using {\it SOHO/EIT}
\citep{krucker1998,bk2002} and {\it TRACE} \citep{parnell2000,asch2000}
providing energies between $10^{24}$ to $10^{27}$~erg. Small events in
active regions, termed ``active region transient brightenings'', were seen
in soft X-rays with {\it Yohkoh/SXT} \citep{shimizu1995}, with energies
between $10^{26}$ to $10^{29}$~erg. All of these studies found the power
law index of the frequency distributions to be between $\alpha=1.5-2.6$.
\citet{asch_parnell2002} investigated the effect of instrumental bias on
the different indices from {\it SOHO/EIT}, {\it TRACE} and {\it Yohkoh/SXT}
data. \citet{parnell2004} later pointed out that the discrepancy between
power-law indices from different instruments can be due to the effects of
least-squares fitting some the of binned histograms. Similar indices were
obtained when a maximum likelihood method \citep{parnell2000} was
used instead.

\begin{figure*}\centering \includegraphics[scale=0.8]{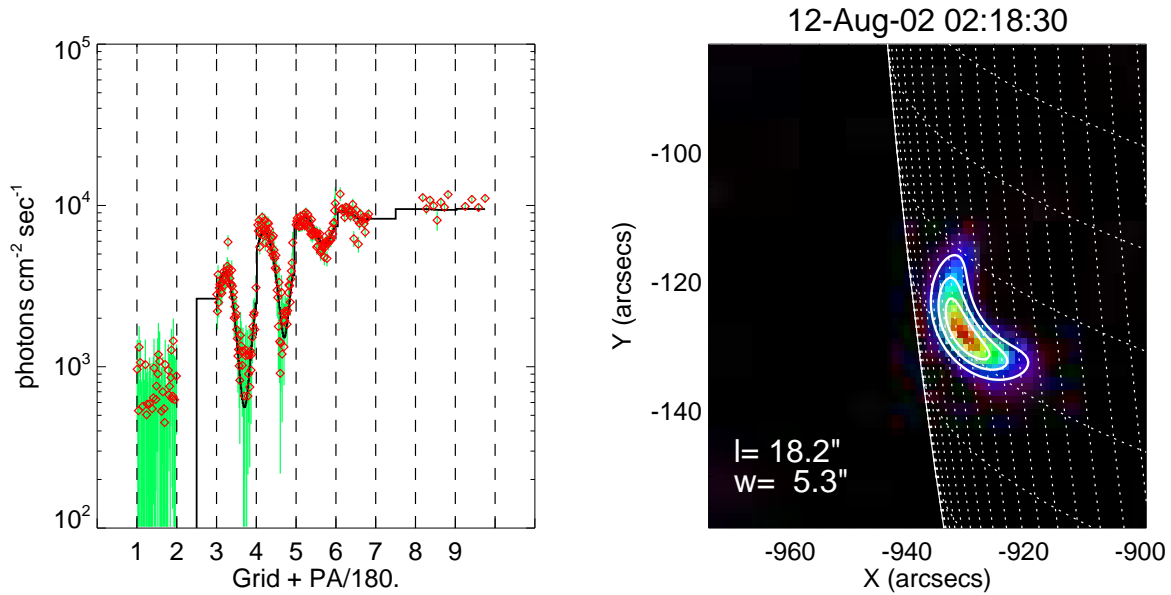}
\caption{\emph{(Left)} Visibility forward fit model loop shape (solid line)
onto the 4-8~keV visibilities amplitudes (diamonds with error bars) as a
function of the phase for each grid. The amplitudes for each
grid/subcollimator are shown between the dashed vertical lines, indicating
the amplitudes measured as the grids rotate from a position angle of
$0^\circ$ to $180^\circ$. Thus an amplitude measured in subcollimator 3
when rotated to a position angle of $45^\circ$ would be plotted against
3.25 on the x-axis. \emph{(Right)} Resulting image of the visibilities shown
in left panel, background image found using MEM NJIT algorithm
\citep{schmahl2007}, overplotted contours (25 \%, 50 \% and 75 \%)
representing the model shape. The loop length $l$ and width $w$ are
quoted. Color version available in electronic edition.} \label{fig:visfitex}
\end{figure*}

The non-thermal hard X-ray emission is assumed to be due to a power-law
distribution of electrons emitting hard X-rays via bremsstrahlung in a thick
target. The resulting power-law photon spectrum reflects this electron
distribution \citep{brown1971} allowing the power in these accelerated
electrons above a low-energy cut-off $E_\mr{C}$ (in keV), to be calculated
as

\begin{eqnarray}\label{eq:pow}
P_\mr{N}(\ge E_\mr{C})&=&9.5\times 10^{24} \gamma^2 (\gamma-1) \nonumber\\
&&\times \beta \left(\gamma -\case{1}{2},\case{3}{2} \right)
I_\mr{0}E_\mr{C}^{(1-\gamma)} \quad \mr{erg~s}^{-1} \end{eqnarray}

\noindent where $\gamma$ and $I_\mr{0}$ are the index and
normalization of the photon power-law spectrum (in units of photon flux,
s$^{-1}$ cm$^{-2}$ keV$^{-1}$) and $\beta(m,n)$ is the beta function
\citep{brown1971,lin1974}. Therefore observing the hard X-ray spectrum of
these events for various time intervals during each flare is sufficient to
obtain an estimate of the non-thermal energy. However there is ambiguity
in the low-energy cut-off, $E_\mr{C}$, because the observed photon
spectrum depends only weakly on it, with a resulting flattening of the
photon spectrum below $\epsilon_\mr{B}$ not uniquely related to
$E_\mr{C}$, with $\epsilon_\mr{B} \leq E_\mr{C}$ \citep{holman2003}.
Uncertainty in $E_\mr{C}$ results in a larger uncertainty in the power
estimate: a factor of 2 increase/decrease in $E_\mr{C}$ would result in a
factor of 8 decrease/increase in the power for flat spectra ($\gamma=4$)
or a factor of 64 decrease/increase in the power for steep spectra
($\gamma=7$).

Previous statistical studies of the non-thermal energy in flares used the
energy threshold of the instrument as an estimate of $E_\mr{C}$, as they
did not observe to low enough energies nor had sufficient energy
resolution to observe the flattening of the spectrum. \citet{crosby1993}
using {\it SMM/HXRBS} estimated the non-thermal energy in large flares
$>25$~keV, finding that the power distribution at the peak time of
emission had a power-law with $\alpha=1.67$ over $10^{27}$ to
$10^{30}$~erg~s$^{-1}$ and the total non-thermal energy of these events
had $\alpha=1.53$ over $10^{27}$ to $10^{32}$~erg. Microflares were
observed down to 8~keV with {\it CGRO/BATSE} \citep{lin2001}, finding
energies over $10^{27}$ to $10^{30}$~erg, and by
\emph{WATCH/GRANAT} down to 10~keV \citep{crosby1998}.

The Reuven Ramaty High Energy Solar Spectroscopic Imager {\it RHESSI}
\citep{lin2002} is uniquely suited to investigate both the energy in the
heated and accelerated electron populations, thermal and non-thermal
emission, of these small events. This is due to its unprecedented
sensitivity to 3-25~keV X-rays, high spectral resolution and imaging
capabilities \citep{krucker2002}. Previous studies of {\it RHESSI}
microflares have concentrated on individual events
\citep{krucker2002,benz2002,benz2003} or small samples, often compared
to other wavelengths,
\citep{liu2004,qiu2004,battaglia2005,kundu2005,kundu2006,stoiser2007}.

Here we present the first analysis of all {\it RHESSI} microflares found as
transient bursts in 6-12~keV during periods of shutter-out observations, as
detailed in part I of this article \citep{mfpart1}. Between March 2002 and
March 2007 25,705 events were found. These are active-region
phenomena of low C {\it GOES} class to below A-class. For the analysis
presented here we measure the energy for 16~seconds around the time of
peak emission in 6-12~keV. The peak time is only used as it presents the
best opportunity to obtain enough counts above background to permit
analysis in each event. Given the standard flare time profile of a sharp
impulsive rise followed by slower decay phase, this means that this
analysis will mainly cover the impulsive phase of these microflares.

The calculation of the thermal energy using equation (\ref{eq:therm})
requires knowledge of the volume of the thermal source and thermal
parameters from the spectrum. Imaging these events, and hence
estimating the volume of thermal emission, is detailed in \S\ref{sec:vis}. In
\S\ref{sec:ospex} the spectral fitting of the events is described, obtaining
the thermal and non-thermal parameters of these events. The relationship
between these parameters and those found using {\it GOES} data are
discussed in \S\ref{sec:cor}. The calculation of the thermal energy and
power in non-thermal electrons and the resulting frequency distributions is
presented in \S\ref{sec:eng}. We discuss these results further, including
how this analysis at peak time relates to the emission over the whole of
the microflare, in \S\ref{sec:discons}.

\section{Imaging Using Visibilities}\label{sec:vis}

{\it RHESSI} imaging is achieved through a Fourier-based method using
rotation modulation collimators (RMCs) \citep{hurford2002}.  Each RMC
time-modulates sources whose size scale is smaller than its resolution.
This spatial information encoded in the time-modulation profile is normally
reconstructed into an image via techniques such as back projection
\citep{hurford2002}.

A recently implemented alternative technique converts the
time-modulation profile to complex visibilities before recovering the
spatial information \citep{hurfordvis}. Each {\it RHESSI} visibility is a
calibrated measurement of a single Fourier component of the source
distribution measured at a specific spatial frequency, energy and time
range. The visibilities are the complex quantities obtained from the fitted
amplitude and phase of the modulated time profile for a particular roll
orientation (rotational phase) of an individual RMC. The resulting set of
visibilities for all roll angles and RMCs is a calibrated and compact
representation of the original time profile, with little loss of information.

The advantage of using visibilities is twofold. First, as a smaller data set
has to be processed, the  image reconstruction from the visibilities is
considerably quicker than using the time profile directly. Second, the
visibilities are fully calibrated measurements, representing an
intermediate step between the modulation profile and imaging, meaning
that spatial information can be found directly from the visibilities without
having to compute an image. This has been implemented in a Visibility
Forward Fit (VFF) algorithm \citep{hurfordvis} which determines the best-fit
parameters, with statistical errors, for simple assumed source geometries
(elliptical Gaussian, curved elliptical Gaussian, multiple sources etc).

We are primarily interested in the size of the thermal source, in order to
make an estimate of the density and hence thermal energy, see
\S\ref{sec:eng}. The images of thermal sources in microflares are taken
over 4-8~keV and generally have a single elliptical source or loop shape.
We fit a 2D model of a curved elliptical Gaussian to the 4-8~keV visibilities
for 16~secs about the peak in 6-12~keV for each microflare. This attempts
to fit a Gaussian profile along the curved semi-major axis, equivalent to
the loop arc length. If the source is not appreciably curved an elliptical
geometry of zero curvature is returned. Seven parameters are obtained
from each fit of this 2D model: the centroid position (x,y), photon flux,
FWHM loop length and width, curvature and position angle of the
semi-major.

\begin{figure*}\centering \includegraphics[scale=0.5]{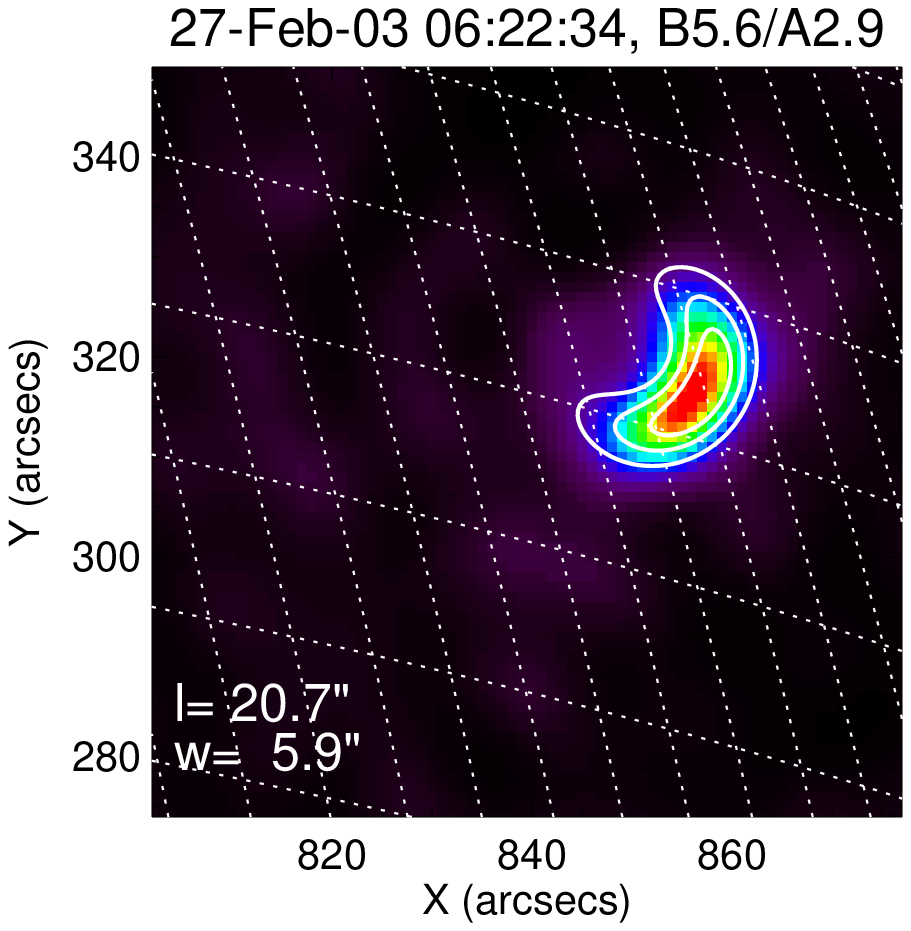}
\includegraphics[scale=0.5]{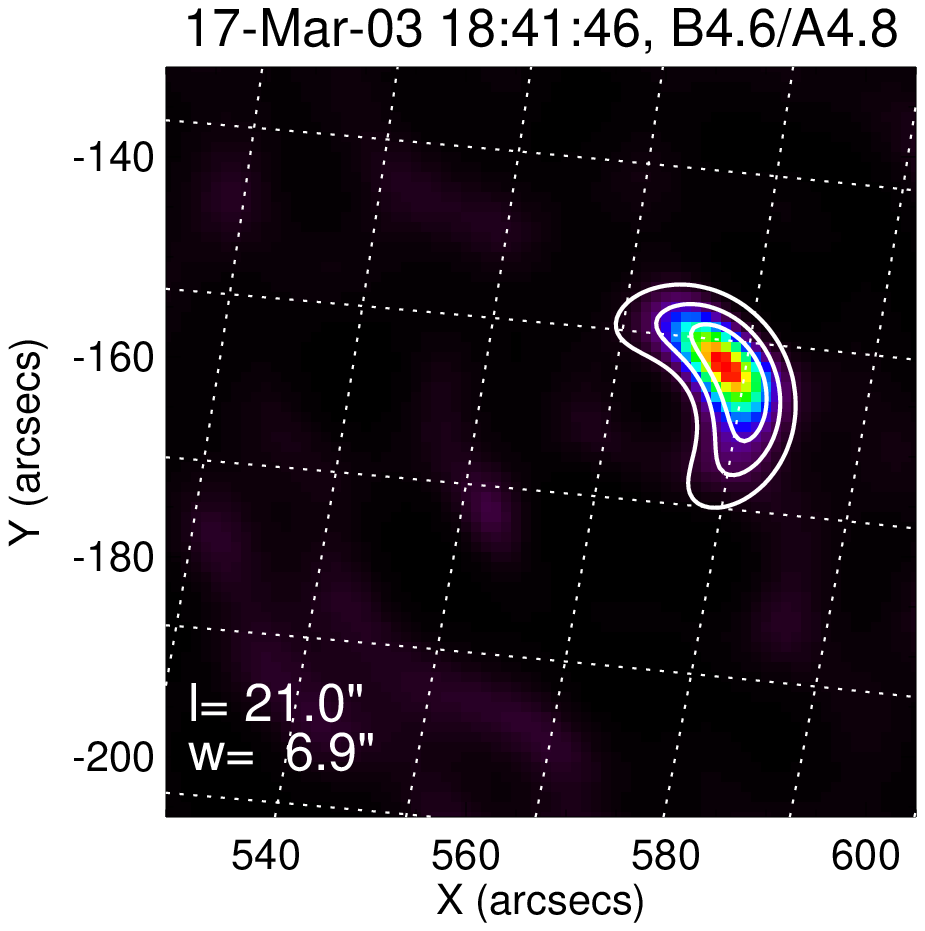}
\includegraphics[scale=0.5]{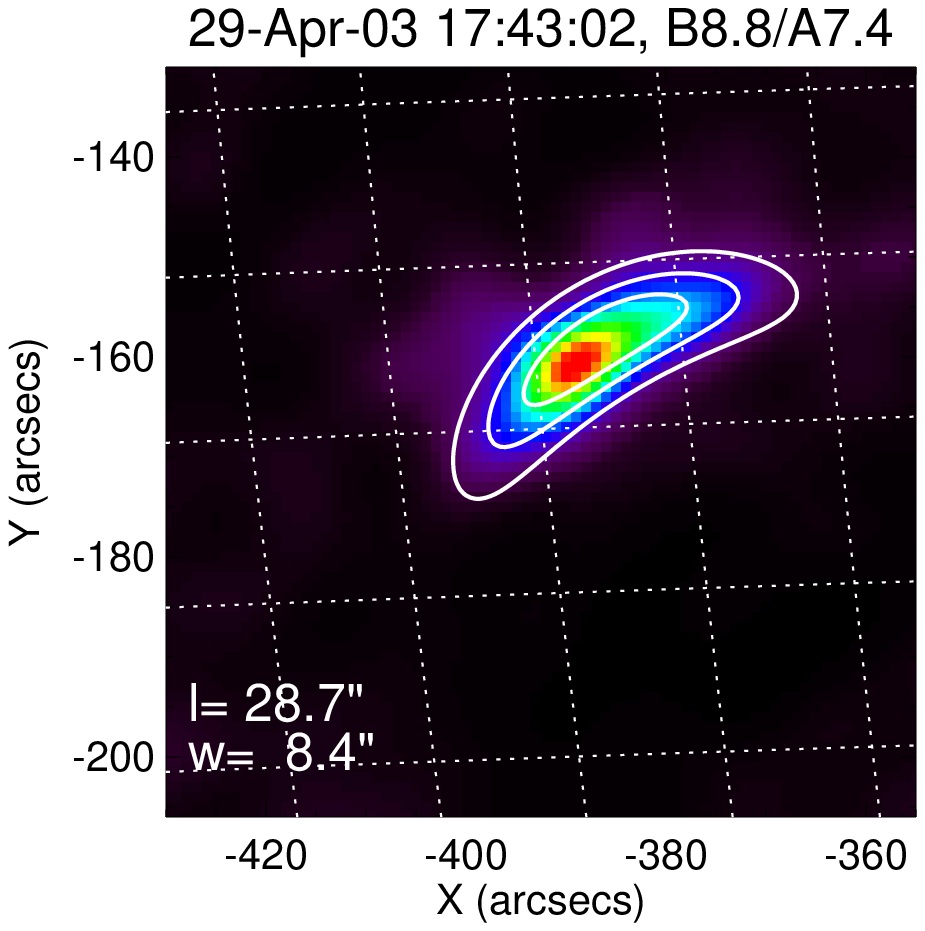}\\
\includegraphics[scale=0.5]{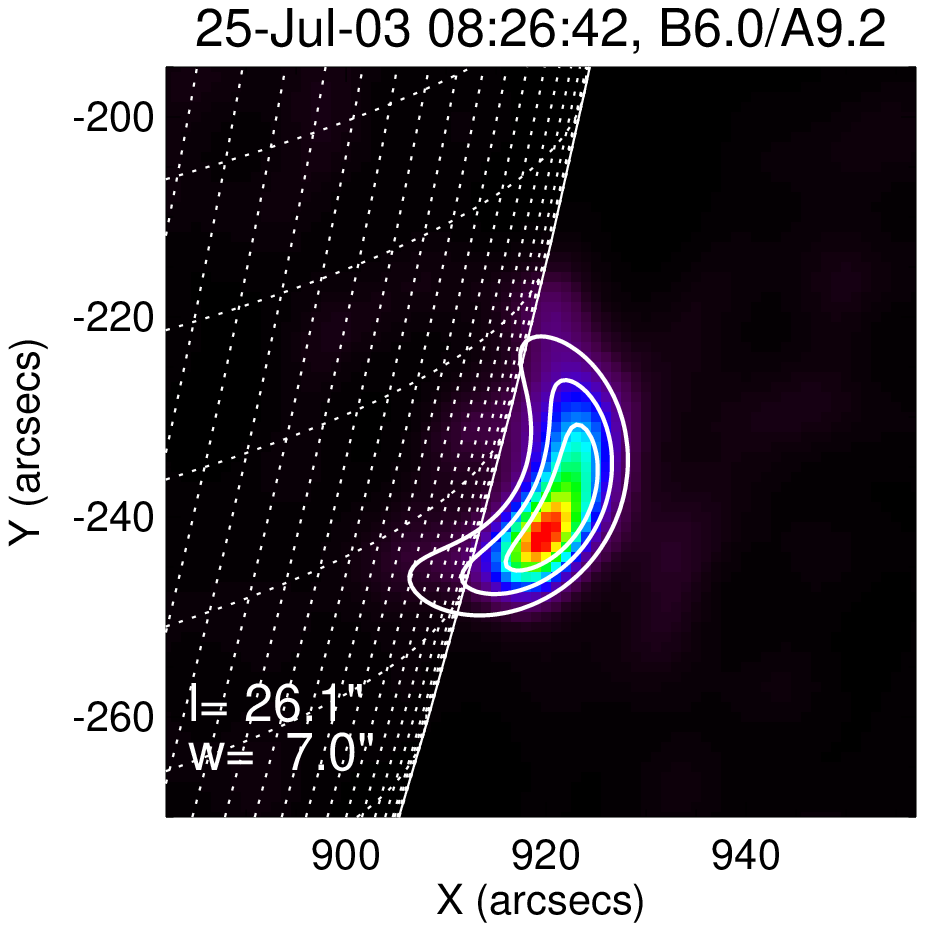}
\includegraphics[scale=0.5]{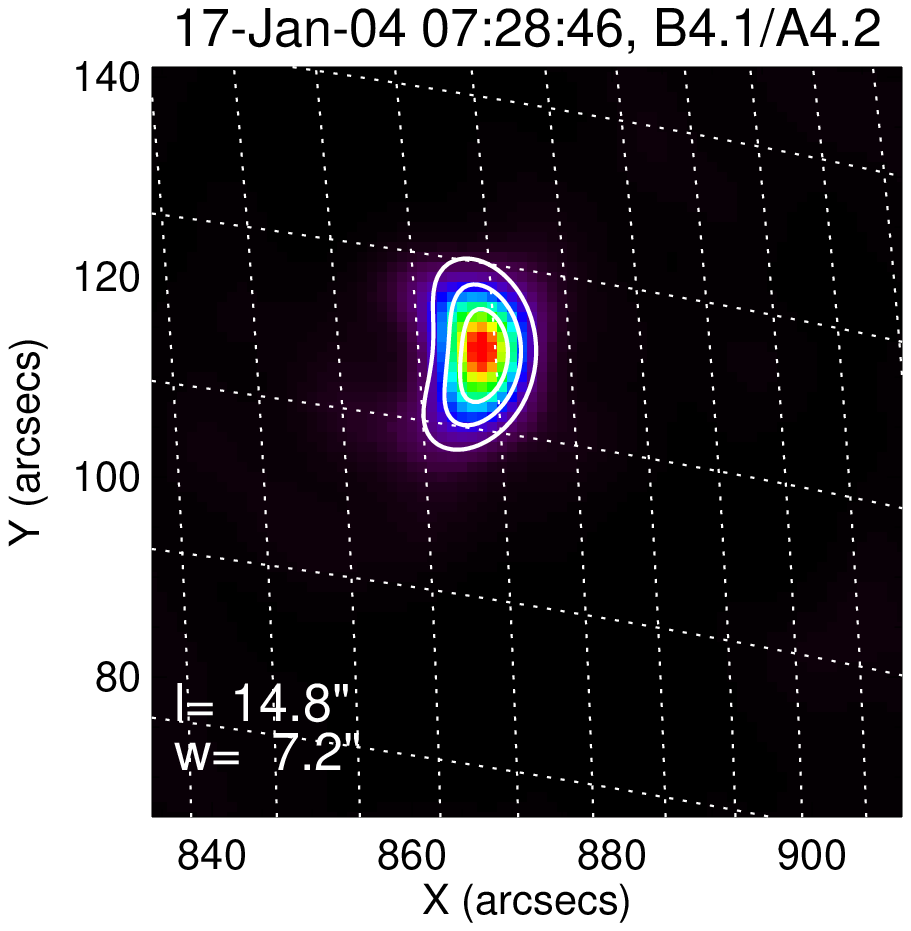}
\includegraphics[scale=0.5]{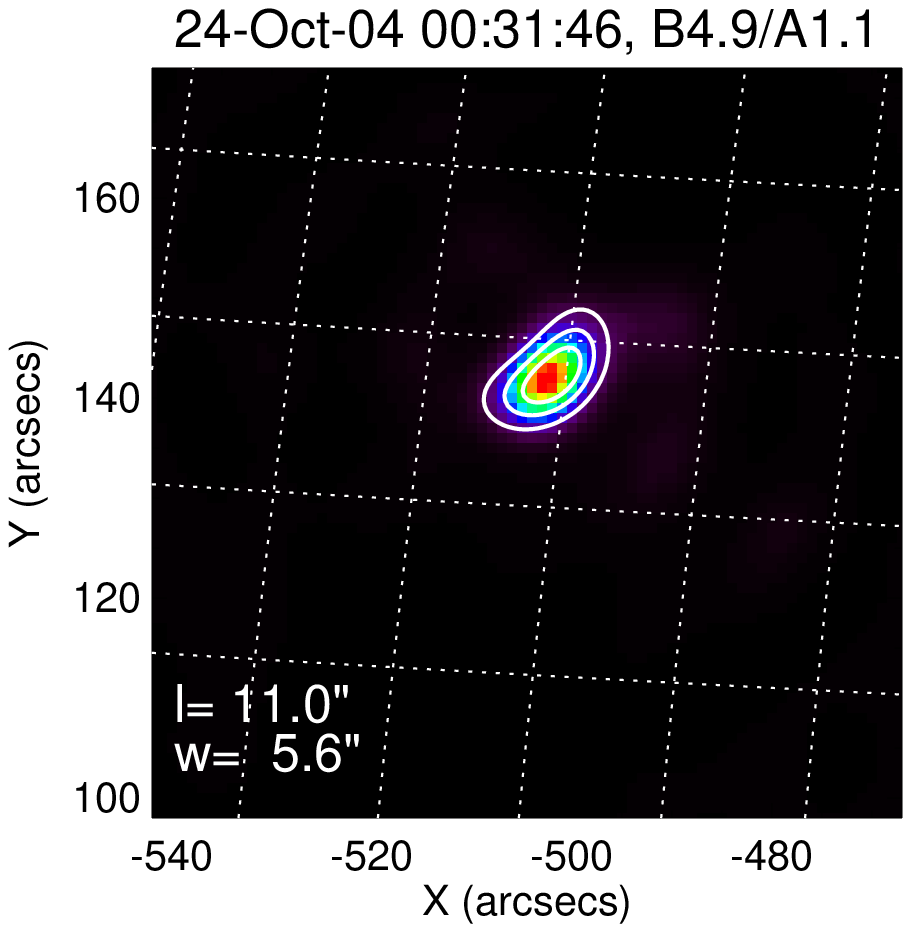}\\
\caption{Example microflares showing the MEM NJIT image (background)
and the forward fit model shape (foreground contours of 25\%, 50\% and
75\%). The model loop length $l$ and width $w$ are quoted in each panel.
Color version available in electronic edition.}\label{fig:imgexam}
\end{figure*}

An example of this fitting is shown in Figure \ref{fig:visfitex}. Here in the
left panel we have the visibility amplitudes,  with statistical errors, for
each grid and is position angle. The solid line is the VFF model loop shape
which has fitted subcollimators 3,4,5,6,8 and 9 well. For this event the
finest subcollimator, 1, is dominated by noise and so has little influence on
the fit. Subcollimators 2 and 7 are not included as they provide a poor
response in this energy range. The right panel shows the resulting image
of these visibilities, produced using the MEM NJIT algorithm
\citep{schmahl2007}, with the VFF model loop from the left panel
overplotted as contours.  The contours correspond well with the
background image. Further examples of the resulting fits are shown in
Figure \ref{fig:imgexam}, again calculated using detectors 1,3,4,5,6,8 and
9.

\begin{figure}\centering
\plotone{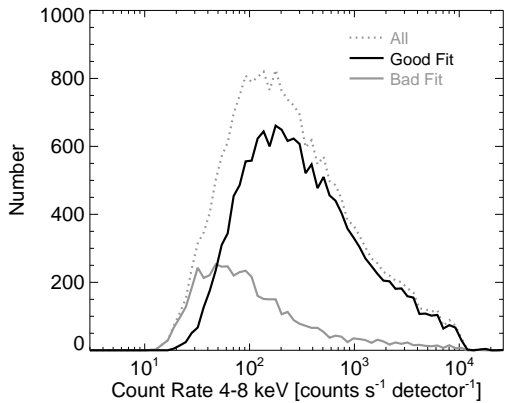}\\
\caption{Histogram of the count rate per detector in 4-8~keV, the different lines
indicating all microflares (light grey dotted) and subsets with a good
(solid black) and bad (solid grey) model fits to the
visibilities.}\label{fig:histrate48} \end{figure}

The model fit for each of the microflares not only provides the spatial
information but also a measure of the ``quality'' of the fit, based upon
whether the fit converged, whether the fit parameters reach the limit of
their range and the size of the errors relative to the parameter. Such an
objective measure of the fit quality is vitally important for an automated
analysis project as it is impractical to visually inspect over 25,000 images.
After processing all microflares we have 18,656 microflares to which the
model achieved a satisfactory fit and were resolved, i.e. returning spatial
sizes larger than the instrumental resolution, 2.3". The majority of events
producing poor fits were those that had the fewest counts. This can be
seen in Figure~\ref{fig:histrate48} where the histogram of the 4-8~keV
count rate per detector of all the microflares is shown, as well as for the
subsets of events producing good and bad fits. There are some microflares
with large count rates but poor fits. This is likely due to an instrumental
issue, such as to an absence of spacecraft roll information.

\begin{figure*}\centering \includegraphics[width=0.32\textwidth]{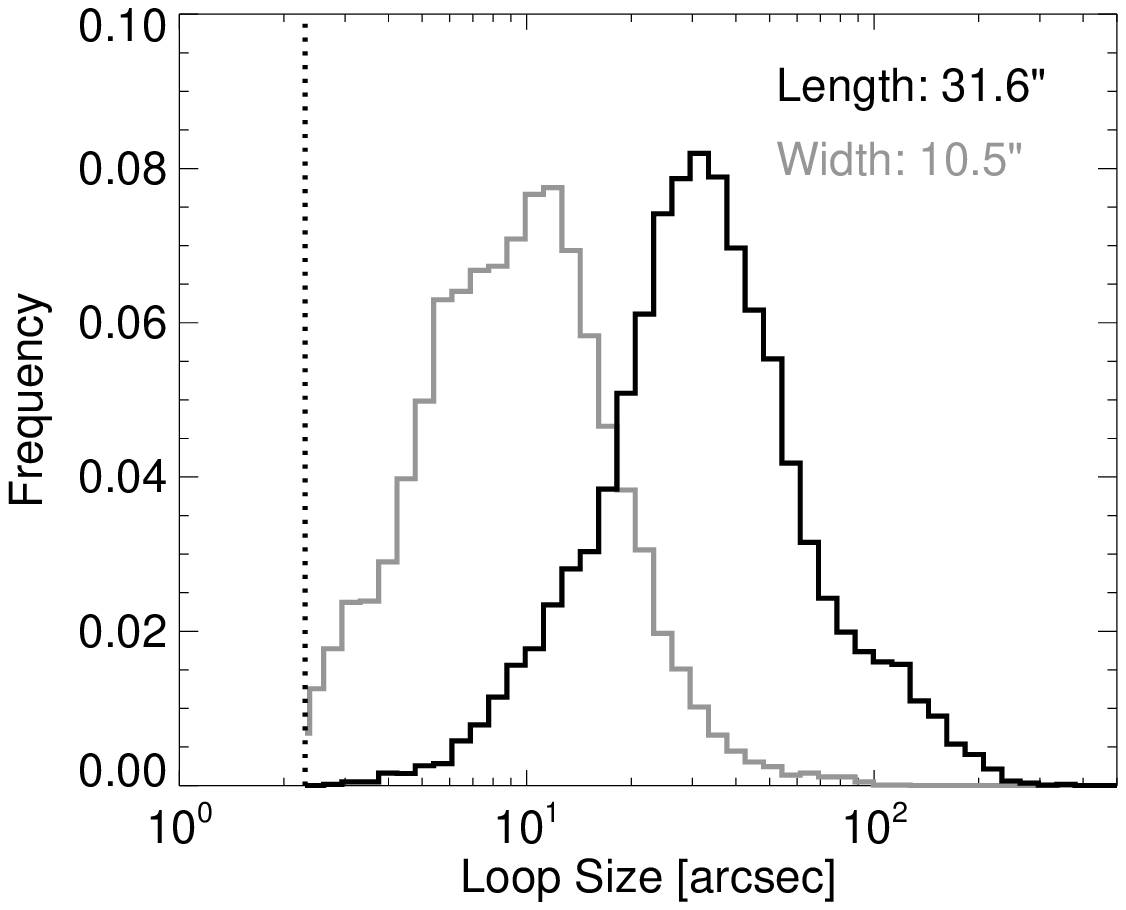}
\includegraphics[width=0.32\textwidth]{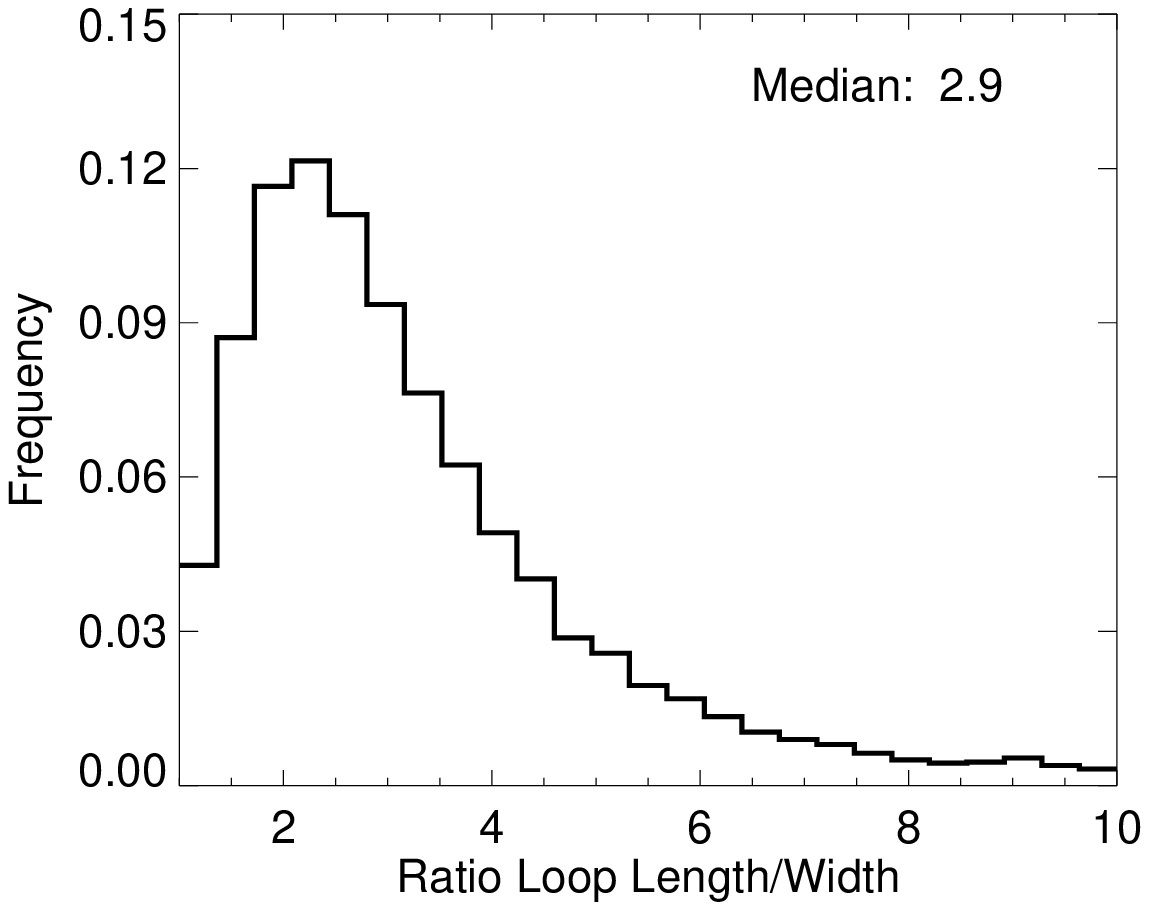}
\includegraphics[width=0.32\textwidth]{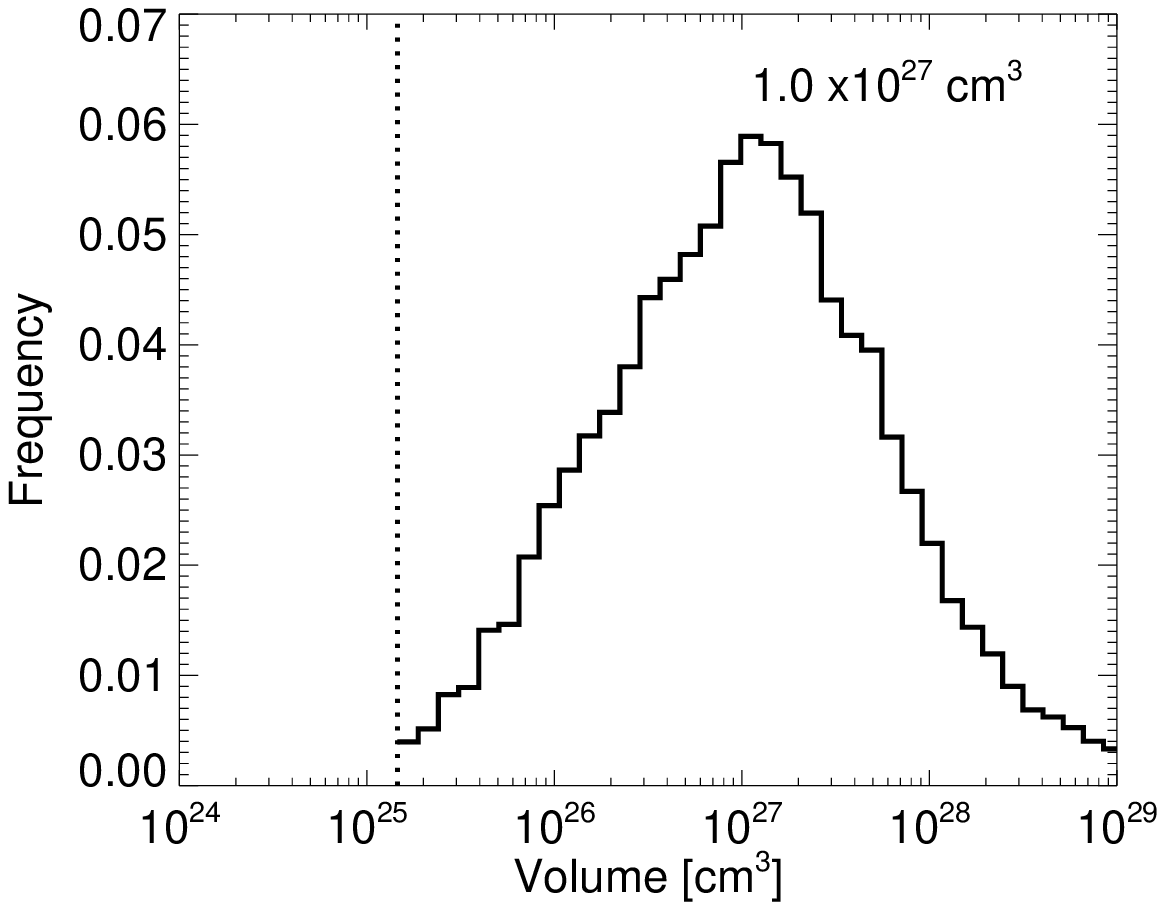}\\
\caption{ (\emph{Left}) Histogram of the model loop arc length and width
for 18,656 {\it RHESSI} microflares. The dotted vertical line indicates the
instrumental resolution limit of 2.3". (\emph{Middle}) Histogram of the
ratio of 4-8~keV microflare loop length to width. (\emph{Right}) Histogram
of the microflare volume of the 4-8~keV loops, assuming cylindrical
geometry from the 2D fitted width and length.  The dotted vertical line
indicates the minimum measurable model volume.}\label{fig:histlwv}
\end{figure*}

\begin{figure}\centering
\includegraphics[scale=0.6]{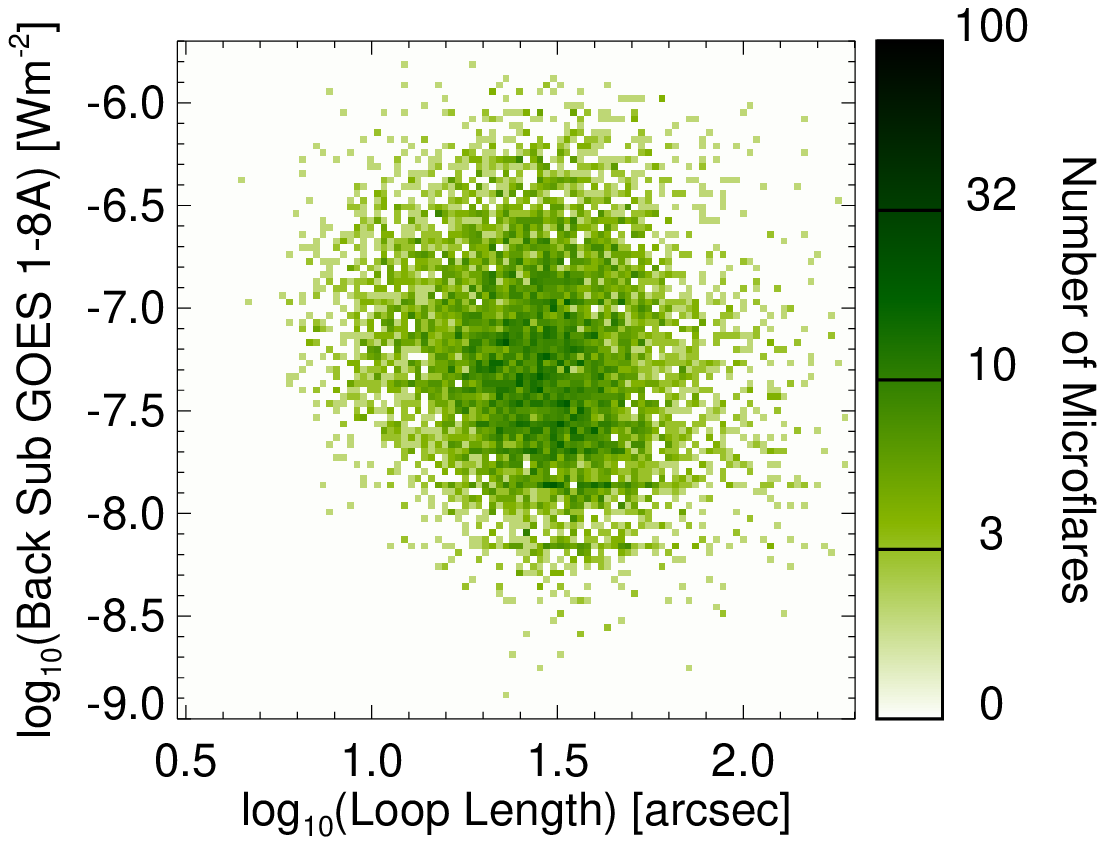}\\
\includegraphics[scale=0.6]{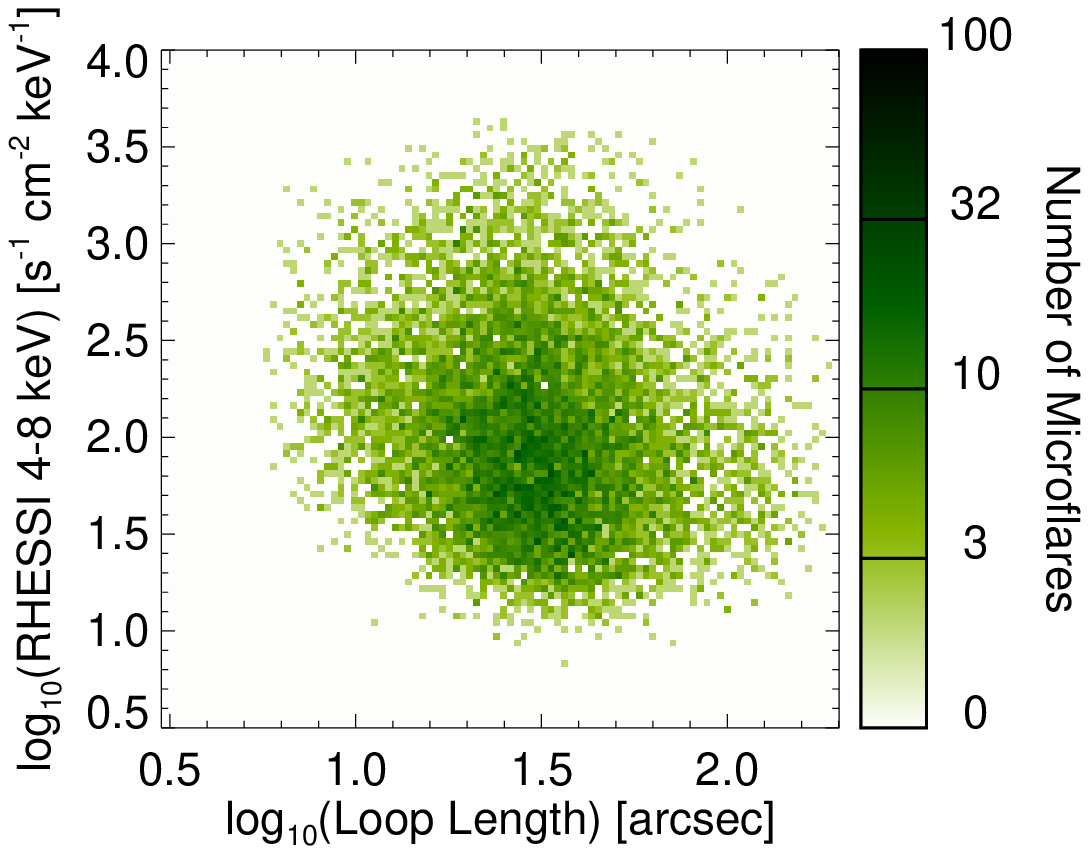}\\
\caption{Correlation plots of the model loop arc length against the background
subtracted GOES 1-8~\AA~soft X-ray flux (\emph{top}) and the {\it RHESSI}
image model's 4-8~keV flux (\emph{bottom}).} \label{fig:lenvsmag}
\end{figure}

The histograms of the loop FWHM arc length $l$ and width $w$ (at loop
mid-point) for the events with good VFFs are shown in Figure
\ref{fig:histlwv}. We find that the median FWHM loop arc length is $31.6"$
(23~Mm) and width is $10.5"$ (8~Mm). The lengths distribution has a
sharp peak, symmetrical in log-space, away from the resolution limit. The
histogram of the ratio of the loop arc length to width (middle panel of
Figure \ref{fig:histlwv}) shows that the majority of the microflare thermal
sources are elongated structures, with the median value of the arc length
being 3 times the loop width. The size of these loops shows no correlation
with the magnitude of the flare, either the flux in the loop or the
background subtracted GOES class (Figure \ref{fig:lenvsmag}).  This shows
that small flares are not necessarily spatially small, which is certainly the
case for the examples shown in Figures \ref{fig:visfitex} and
\ref{fig:imgexam}.

The volume of this thermal emission can be estimated by assuming that
the observed 2D loop structure has a cylindrical geometry as
\begin{equation}\label{eq:vol} V=\pi \left(\frac{w}{2}\right)^2l
\end{equation} \noindent where $l$ is the FWHM loop arc length and $w$
is the width at the loop mid-point. The histogram of the loop volume for
the good events is shown in the right panel in Figure \ref{fig:histlwv}. The
median volume is about $1\times 10^{27}~\mr{cm}^{-3}$, which is a factor
of $\approx$66 larger than the minimum measurable volume of $1.5\times
10^{25}~\mr{cm}^{-3}$, found by taking $w=l=2.3"$ in equation
(\ref{eq:vol}).

\begin{figure}\centering
\plotone{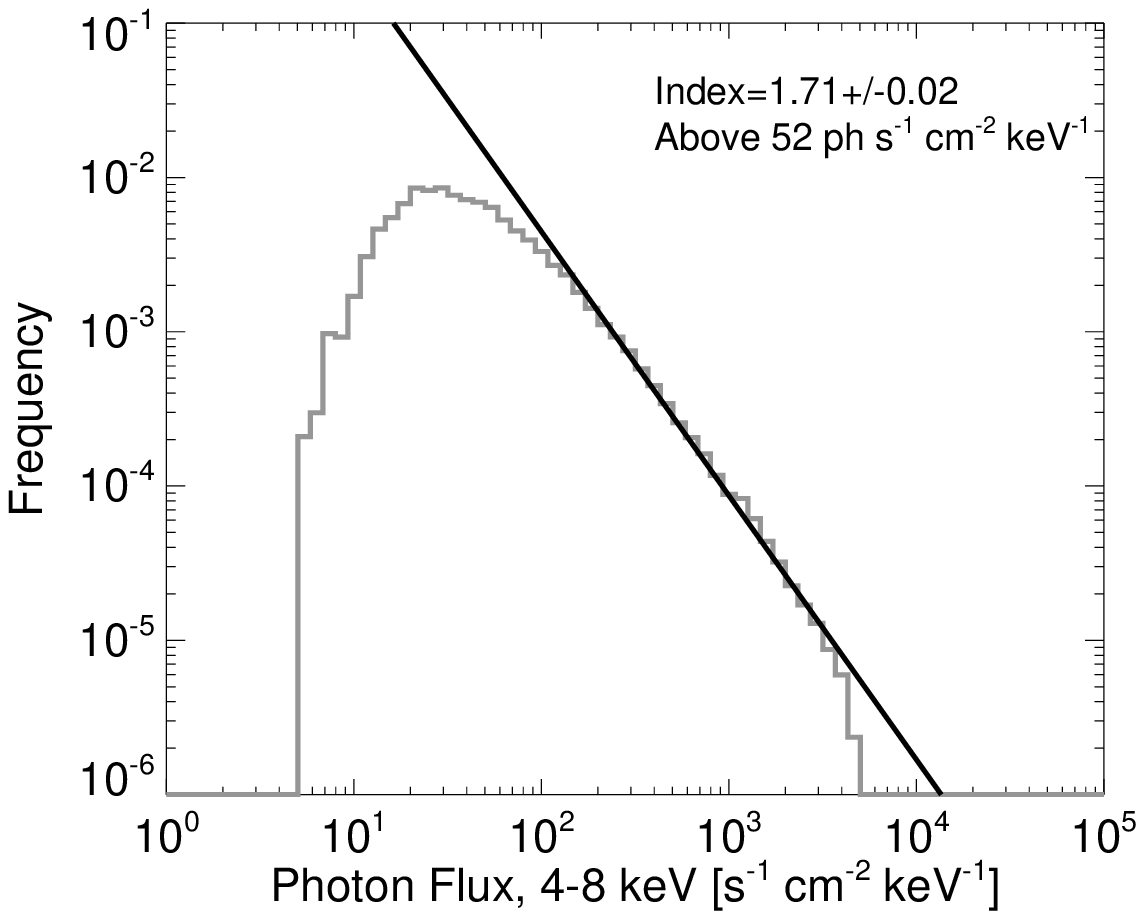}\\
\caption{Frequency distribution of the total photon flux
measured from the VFF 4-8~keV loops. The power-law fit is performed
using a maximum likelihood method of \citet{parnell2000} and is
not a graphical fit to the histogram shown.}\label{fig:histph48}
\end{figure}

We obtain other useful parameters from VFF, in particular a measure of
the total 4-8 keV photon flux from the loop. Since unmodulated background
does not affect the visibilities this flux measure is intrinsically
background-subtracted as it is the emission from only the loop. In Figure
\ref{fig:histph48} we have the differential frequency distribution of this
4-8~keV photon flux. This distribution covers a range of 5 to 5000
photons~s$^{-1}$~cm$^{-2}$~keV$^{-1}$ and power-law parameters  were
found using a maximum likelihood method \citep{parnell2000}. This
technique uses the standard statistical procedure of the maximum
likelihood estimation to fit a skew-Laplace distribution to the data. This
distribution consists of a broken power-law; the index above the break is
the true distribution, whereas below it the power-law fits the
flattening/turning-over of the distribution from under-sampling the
smallest events. So from a simple calculation on the sample (in this case
4-8 keV fluxes) an objective measure is obtained of the power-law index
above a break (with errors found from the 95\% sample confidence)
instead of subjectively choosing bin sizes before line fitting a histogram.
For simplicity the resulting fit is shown overplotted to a standard
histogram in Figure \ref{fig:histph48}. Over two orders of magnitude the
power-law index is $1.71\pm0.02$. This is steeper than the index of
$1.59$ found for hard X-rays $>25$~keV \citep{crosby1993} but flatter
than the that found for soft X-rays, 1.7 -- 2.1
\citep{drake1971,lee1995,feldman1997,veronig2002}. The distribution in
Figure \ref{fig:histph48} deviates from a power-law at both low and high
fluxes due to instrumental selection effects. The events with the smallest
and largest fluxes are missing as we are unable to successfully analyze
these events: the smallest are hard to observe above background and the
largest have excessive counts causing high detector deadtime or are
excluded from our microflare list as {\it RHESSI's} attenuating shutters
were deployed.

\begin{figure}\centering
\plotone{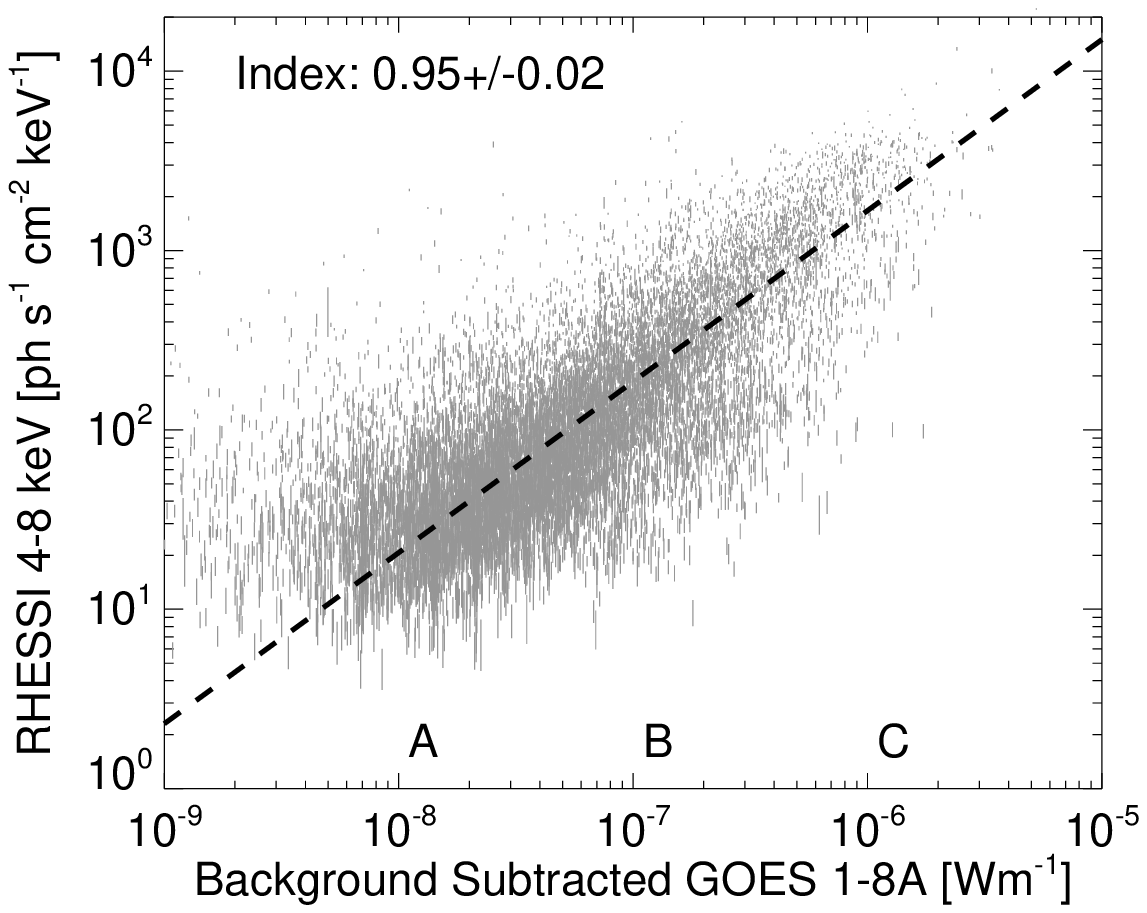}\\
\caption{Background-subtracted
{\it GOES} 1-8~\AA~soft X-ray flux against the {\it
RHESSI} image model's 4-8~keV flux. }\label{fig:goesvsph48}
\end{figure}

\begin{figure}\centering \plotone{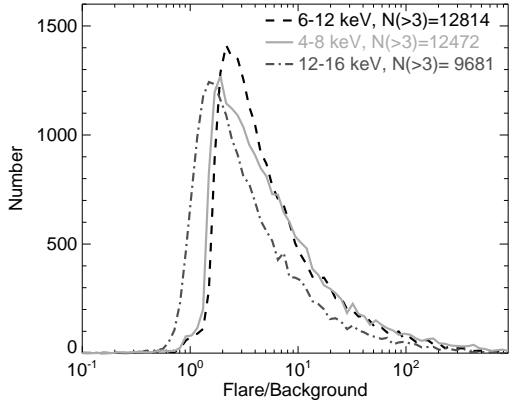} \caption{Histogram of the ratio
of flare signal to pre-flare background in 6-12~keV (dashed black), 4-8~keV
(solid light grey) and 12-16~keV (dashed-dotted grey) at the time of peak
emission in 6-12 keV.}\label{fig:sig2back} \end{figure}

\begin{figure*}\centering \includegraphics[scale=0.5]{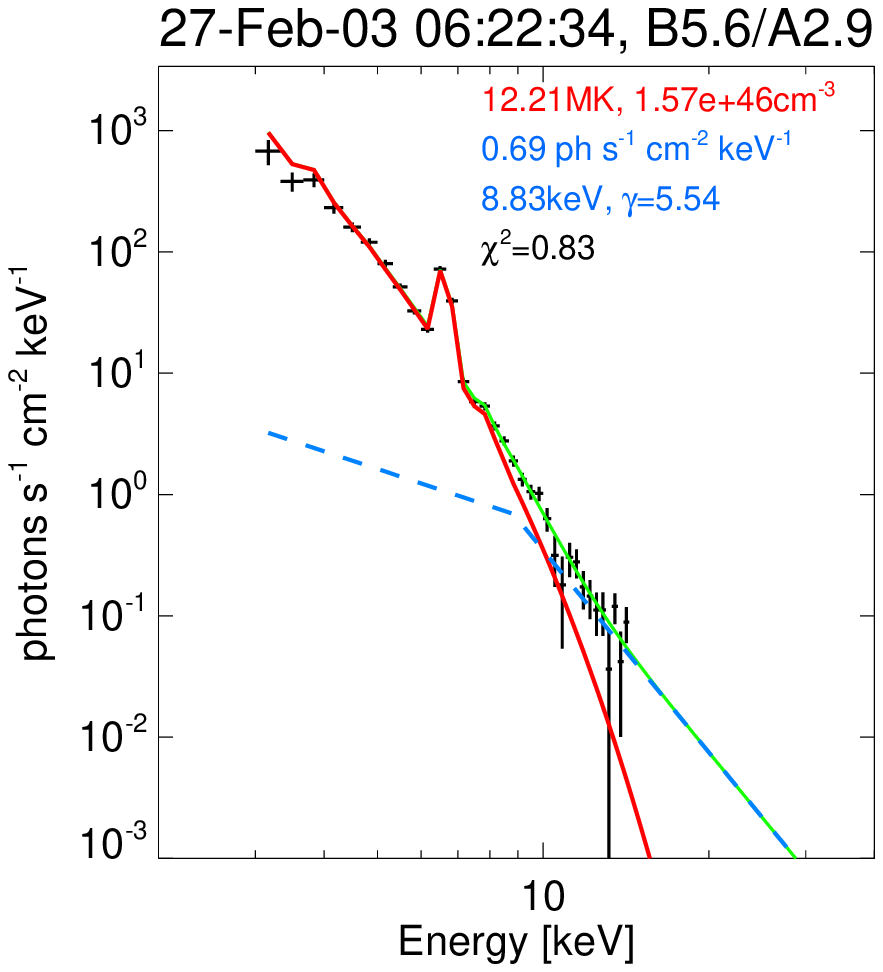}
\includegraphics[scale=0.5]{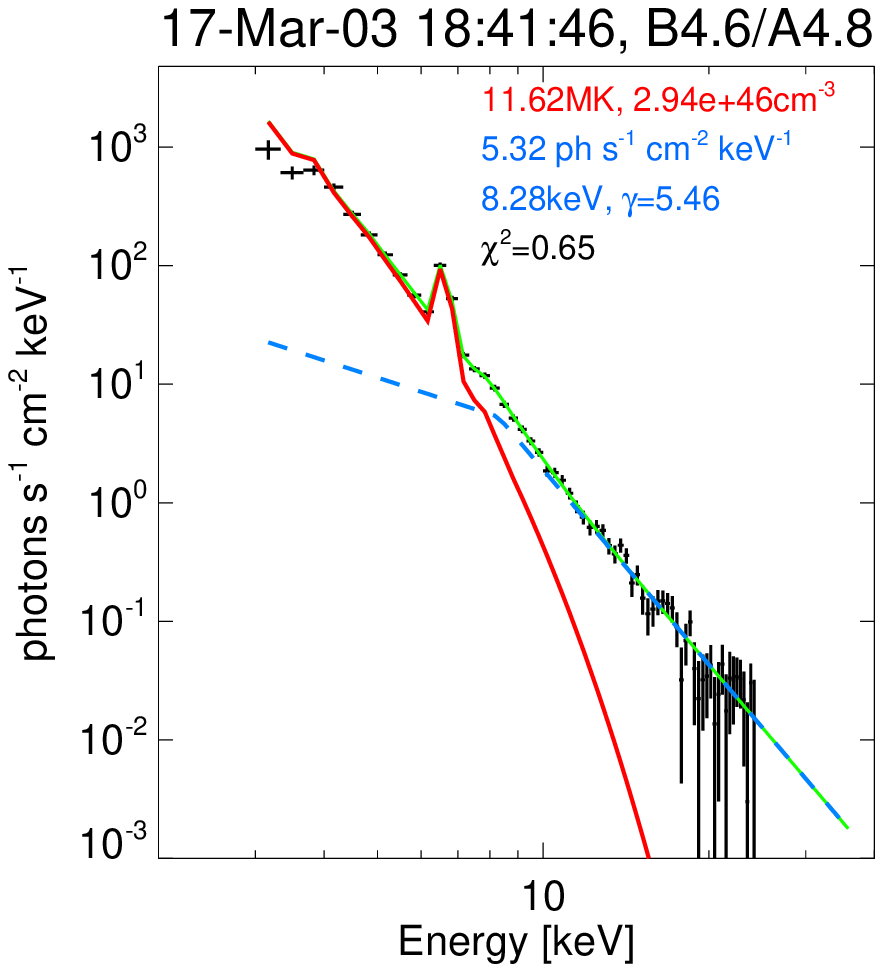}
\includegraphics[scale=0.5]{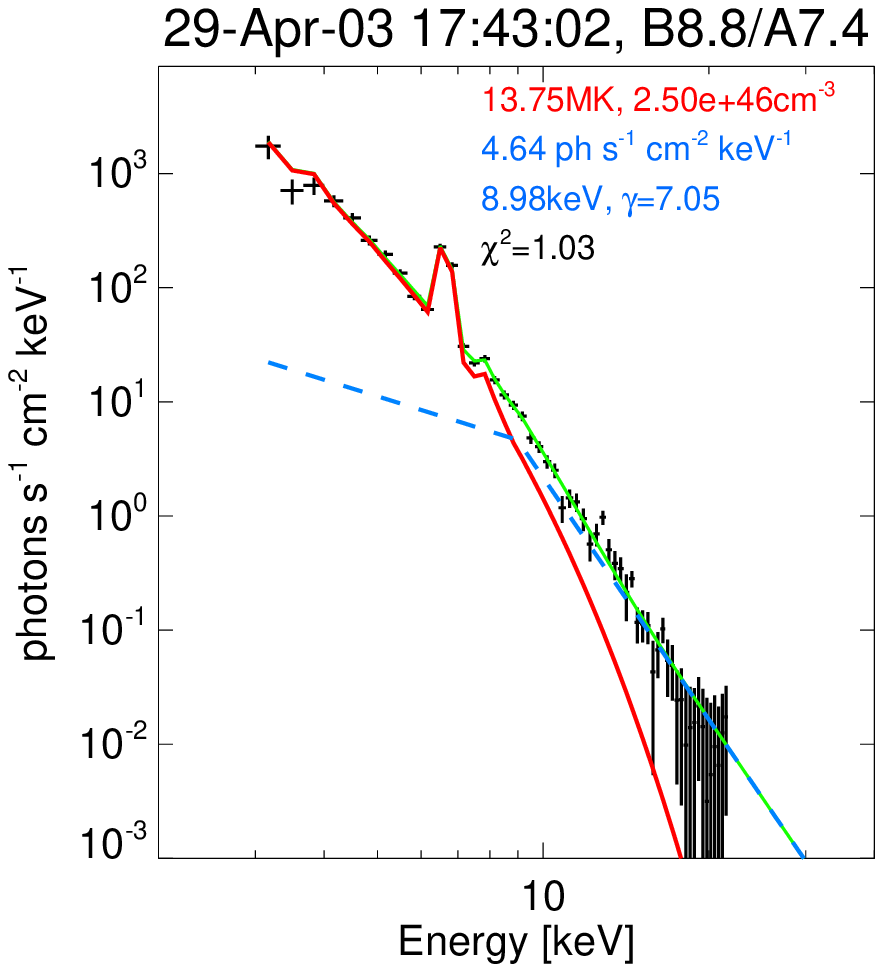}\\
\includegraphics[scale=0.5]{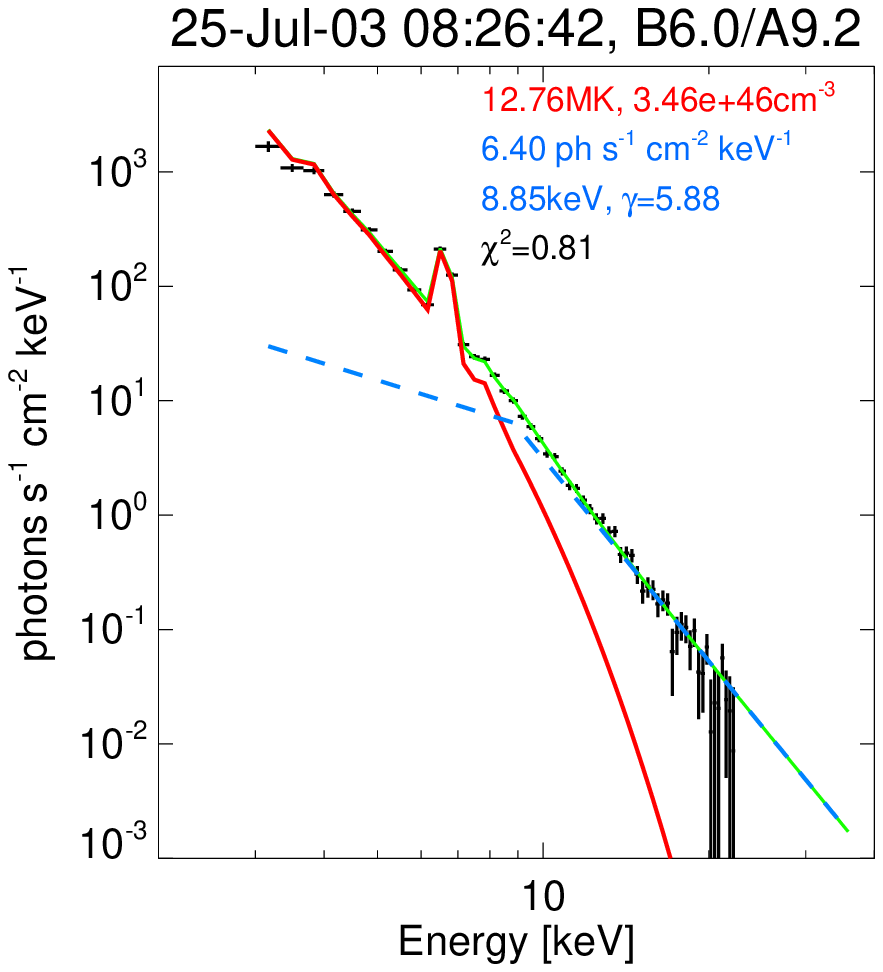} \includegraphics[scale=0.5]{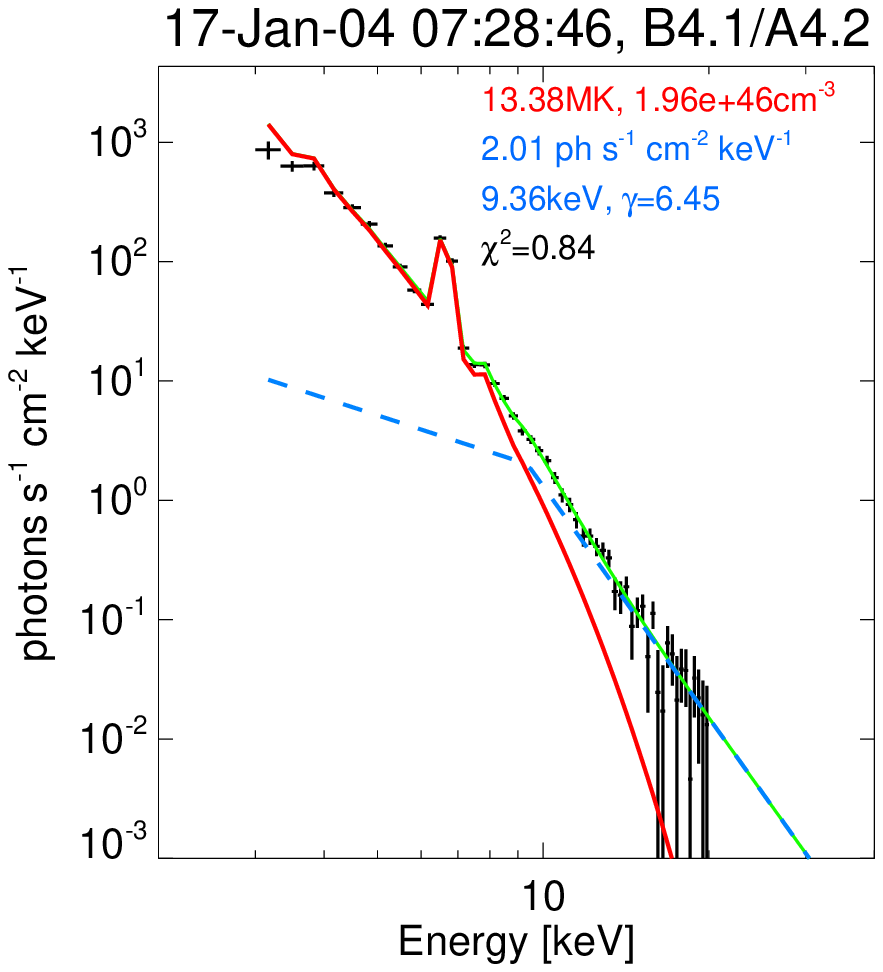}
\includegraphics[scale=0.5]{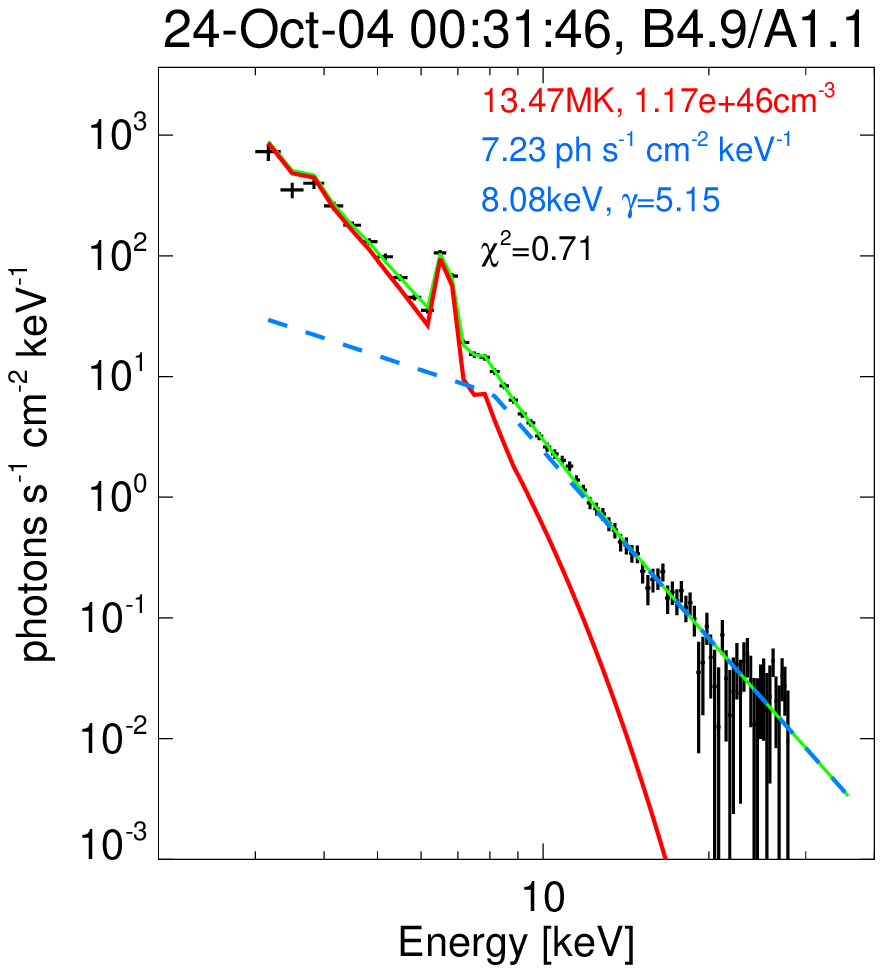}\\
\caption{Example microflare spectra. Shown are the background
subtracted data (black crosses), thermal model (solid line, red in electronic
edition), non-thermal model (dashed line, blue in electronic edition) and
total model (solid line, green in electronic edition). The title shows the {\it
GOES} class and the background-subtracted
equivalent.}\label{fig:specexam} \end{figure*}

Another way in which we can use this {\it RHESSI} 4-8~keV thermal flux is
by comparing it to the emission observed  by {\it GOES} in its 1-8\AA~
band (Figure \ref{fig:goesvsph48}). Here we see that there is a power-law
correlation, with index close to one, although there is some spread about
the fitted line. This suggests that {\it RHESSI} and {\it GOES} are observing
the emission from the same thermal plasma with the different temperature
distribution from event to event accounting for the spread in the
correlation. In comparison to {\it RHESSI} quiet Sun observations
\citep{hannah2007}, the smallest microflare flux measured here is over 2
orders of magnitude larger than the limit found from an active-region-free
quiet Sun. The quiet Sun {\it RHESSI} flux also correlates with the {\it
GOES} flux with a slightly steeper slope ($1.08\pm0.13$) than the
microflares shown here.

\section{Spectrum Fitting}\label{sec:ospex}

The spectrum of each microflare is determined over the same time period
as the visibilities in \S \ref{sec:vis}, 16 seconds around the time of peak
emission in 6-12~keV. These spectra are made with $1/3$~keV energy
bins between 3~keV to 30~keV using detectors 1,3,4,5,8 and 9. For each
microflare, a background interval before and after the event were
determined automatically with the selection of these background times
described in Part I of this paper \citep{mfpart1}. Only the pre-flare
background is used in the subtraction from the spectrum, as it is sharply
defined by the microflare's impulsive phase. This was possible for 19,441
microflares. The histogram of the flare signal to background ratio is shown
in Figure~\ref{fig:sig2back}, for the 6-12~keV, 4-8~keV and 12-16~keV
energy bands. Requiring the flare signal to be at least 3 times the pre-flare
background, we find 12,814 events suitable in 6-12~keV (the energy range
the events were found in) 12,472 in 4-8~keV from predominantly thermal
emission and 9,681 in 12-16~keV from mostly non-thermal emission. This
indicates that the thermal component for more microflares can be
obtained than the non-thermal component. This does not mean that many
microflares do not have a non-thermal component, just that we cannot
distinguish it from the background in these cases.

These background-subtracted observed count spectra are forward-fitted
with a model in photon space converted back to count space using the full
{\it RHESSI} detector response matrix \citep{smith2002} in the OSPEX
software package, an updated version of the SPEX code
\citep{schwartz1996}. The model has both a thermal and non-thermal
component so that we can recover the respective parameters to calculate
the energy in both the heated and accelerated electrons. The thermal
component contains the isothermal bremsstrahlung emission (free-free
and free-bound) as well as line emission from the CHIANTI database
\citep{dere1997,landi2006}. This emission depends on the temperature
$T$ and emission measure $EM=n_\mr{e}^2V$. The non-thermal
component is assumed to be thick-target emission of a power-law
distribution of electrons above a low energy cut-off $E_\mr{C}$, with the
photon spectrum found through numerical integration \citep{holman2003}.
Although this numerical integration provides an accurate representation of
the non-thermal emission it is too slow to compute in this fitting
procedure, requiring multiple iterations per microflare, with tens of
thousands of microflares to fit. Instead the non-thermal component is
fitted with a broken power-law, which has an index of $-\gamma$ above
the break energy of $\epsilon_\mr{B}$ and a fixed index of $-1.5$ below
the break.  An example of this approximation to the numerical integration
is shown in the left panel of Figure~\ref{fig:thick2bpow} and the
relationship between these two models is detailed further in
\S\ref{sec:nonthermpow}, as it is important for calculating the power in the
non-thermal electrons via equation (\ref{eq:pow}).

The fit to each microflare spectrum is conducted using the following
strategy. First the thermal parameters are varied to fit the spectrum over
4-8~keV, where the thermal emission normally dominates. Then the fit is
repeated but the non-thermal parameters are allowed to vary, while
keeping the previously found thermal parameters fixed, to fit the spectrum
from 10~keV up to either 30~keV or to where the background dominates
over the flare signal. The fit is repeated a final time allowing all the fit
parameters to vary to fit the spectrum from 3~keV up to either 30~keV or
to where the background dominates over the flare signal. Examples of
typical microflare spectra and the fits are shown in
Figure~\ref{fig:specexam}. The microflares here illustrate similar
characteristics as seen in previous {\it RHESSI} microflare studies, for
example \citet{krucker2002}. At low energies ($\leq 10$~keV) the thermal
component dominates with the expected spectral lines, Fe K-shell feature
(about 6.7~keV) and the Fe/Ni lines (about 8~keV), for this temperature
range \citep{phillips2004}.  At higher energies ($\ge 10$~keV) there is a
power-law component that dominates over the thermal model which is
normally assumed to be the non-thermal emission. Although this could be
an additional hotter thermal component several other arguments imply
non-thermality: the presence of the Neupert effect \citep{benz2002} in
some cases, imaged hard X-ray footpoints \citep{krucker2002}, and
complementary radio and microwave observations
\citep{liu2004,qiu2004,kundu2005,kundu2006}.

\begin{figure*}\centering \includegraphics[scale=0.475]{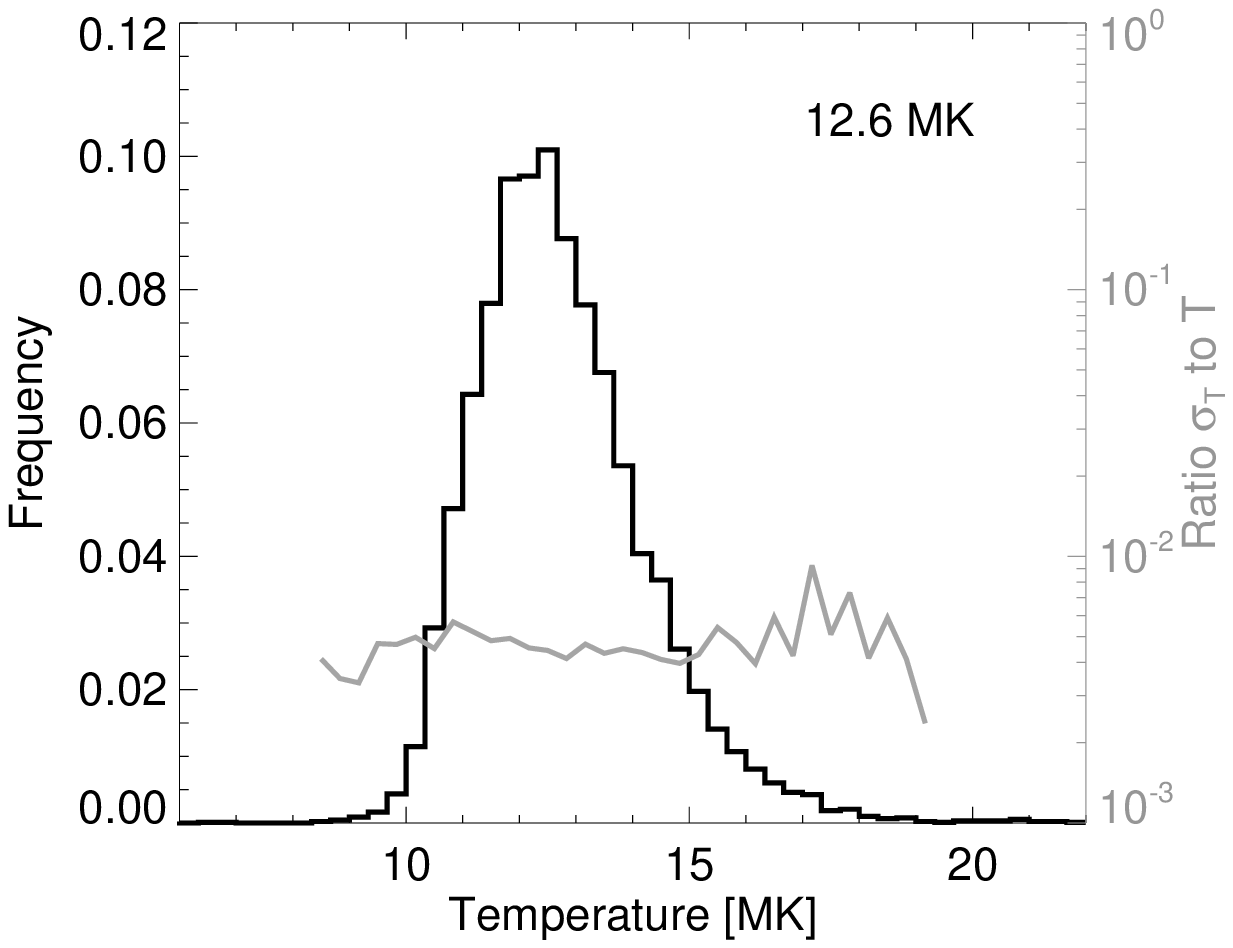}
\includegraphics[scale=0.475]{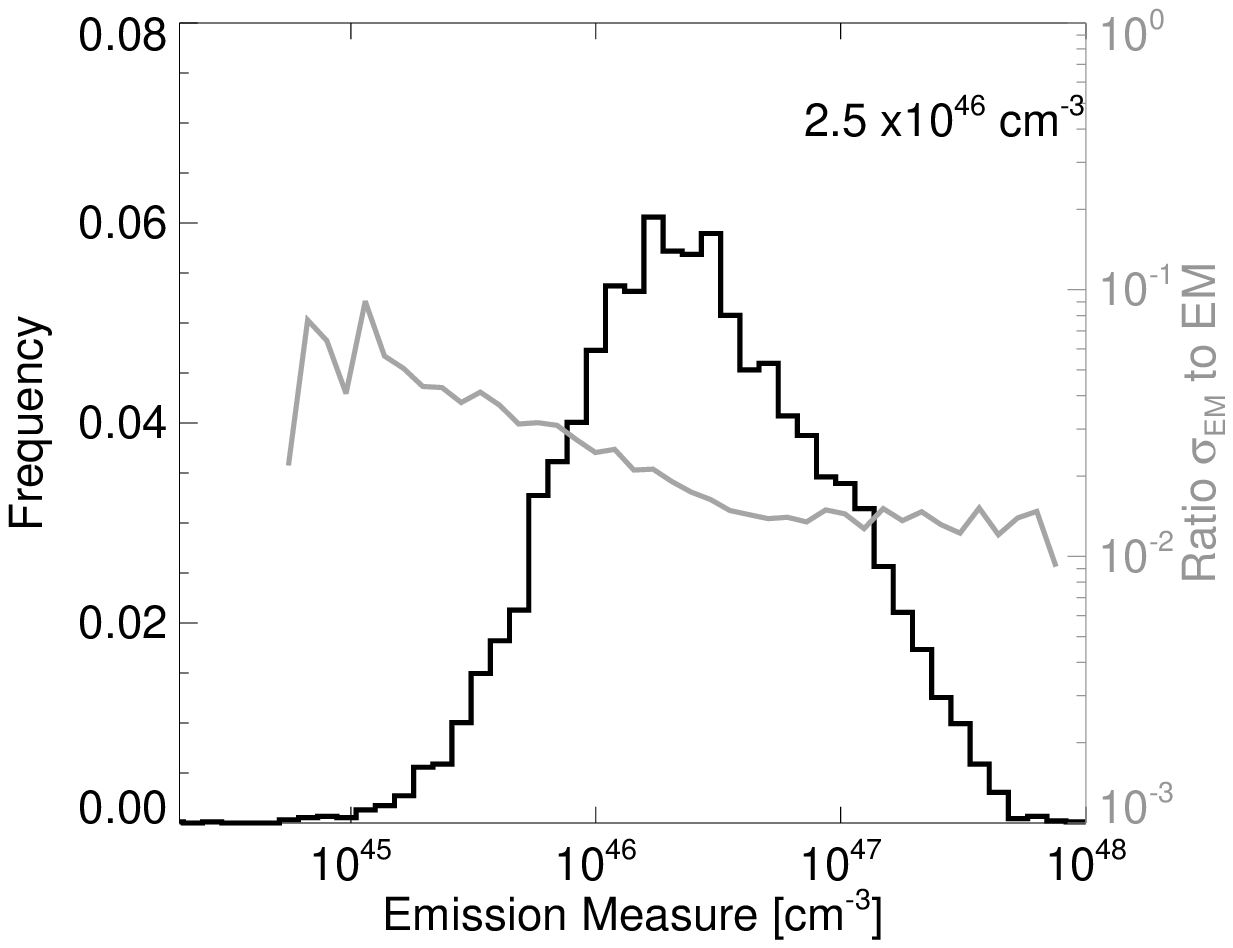}\\
\includegraphics[scale=0.475]{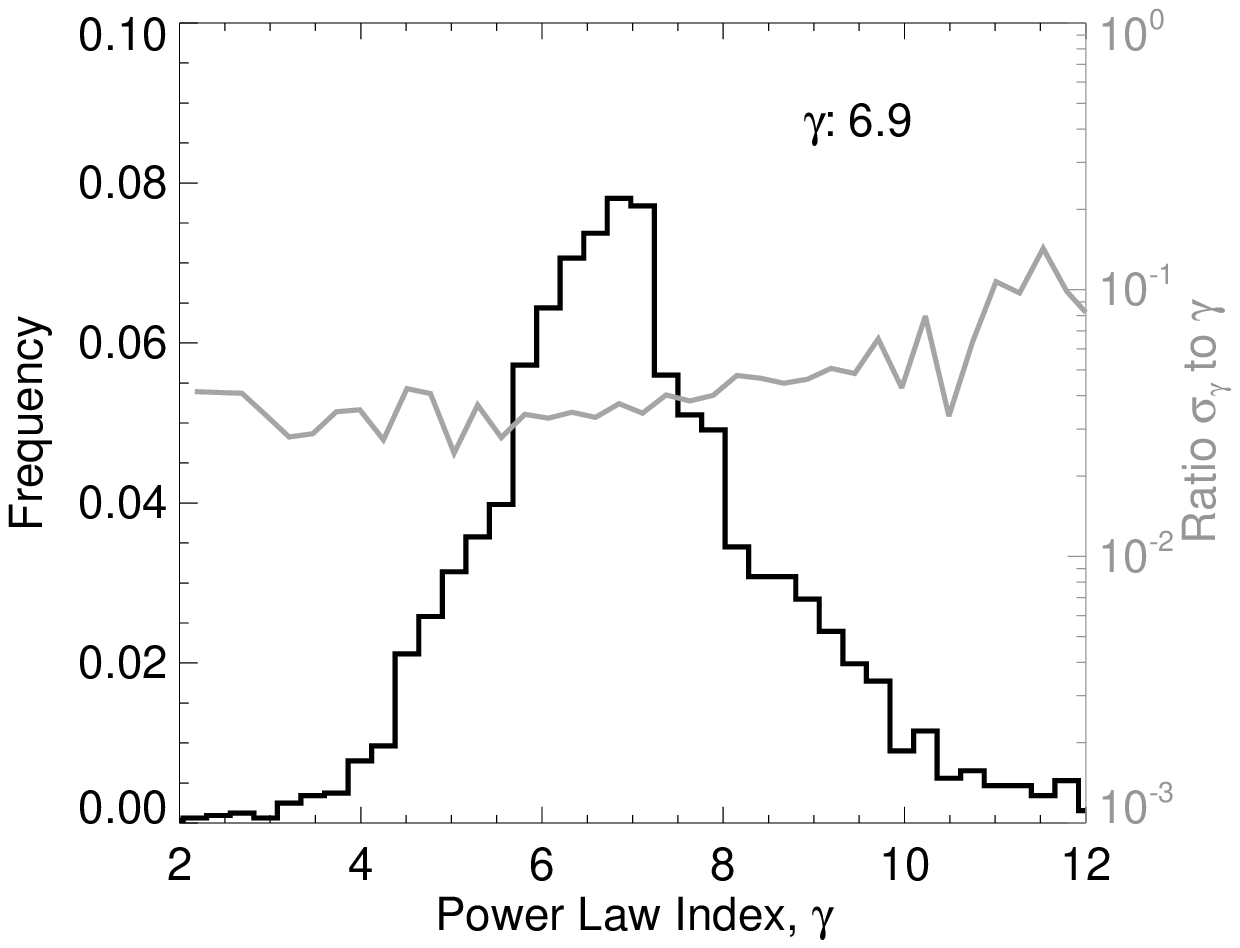}
\includegraphics[scale=0.475]{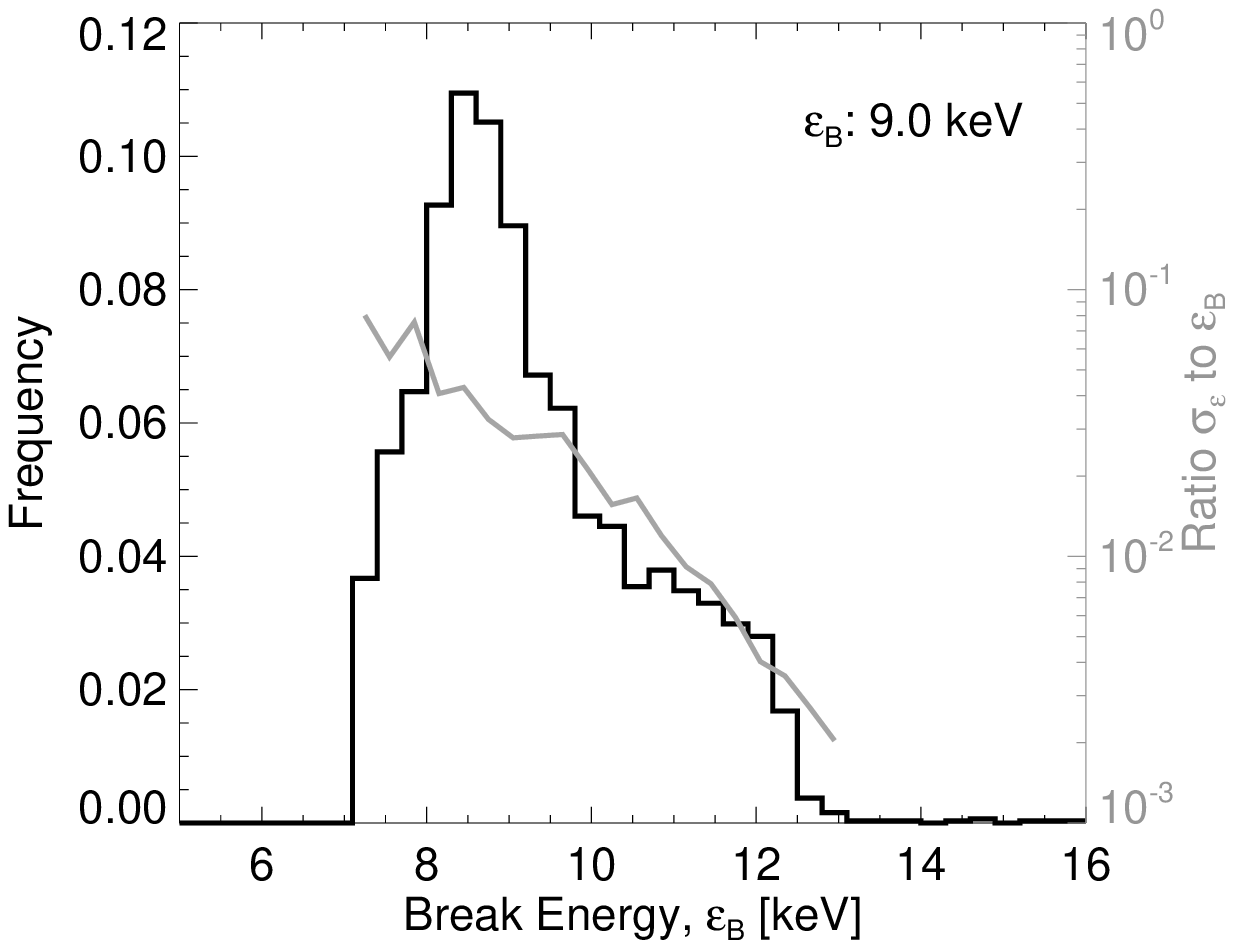}\\
\caption{(\emph{Top}) Histogram of the microflare temperature
(\emph{left}) and emission measure (\emph{right}) from the spectral fits of
9,161 {\it RHESSI} microflares. (\emph{Bottom}) Histogram of the photon
power-law index $\gamma$ (\emph{left}) and break energy
$\epsilon_\mr{B}$ (\emph{right}) for 4,236 {\it RHESSI} microflares. The
grey lines in each panel refer to the righthand axis and show the average
ratio of the error in the fit  to the fitted parameter, as a function of the
parameter.}\label{fig:histtemgeb} \end{figure*}

As with the imaging, the spectral fitting returns parameters that
objectively measure the quality of the fit, such as the fit $\chi^2$ or
whether the fit parameters have large errors or reach the limit of their
chosen fitting range. We obtain 9,161 events for which we trust the fit to
the thermal component of the spectrum and the forward fit model loop.
This is out of a possible 9,693 microflares with good background
subtraction, flare signal-to-background over 4-8~keV and a successful
model fit. For the thermal and non-thermal fit as well as the VFF model
loop fit to the visibilities we obtain 4,236 trustworthy events. This is out of
a possible 8,046 microflares with good background subtraction, flare
signal-to-background over 10-12~keV and successful model fit. The
histograms of these fitted parameters are shown in
Figure~\ref{fig:histtemgeb} and discussed in \S\ref{sec:therm} and
\S\ref{sec:nontherm}.

\subsection{Thermal Parameters}\label{sec:therm}

\begin{figure}\centering
\plotone{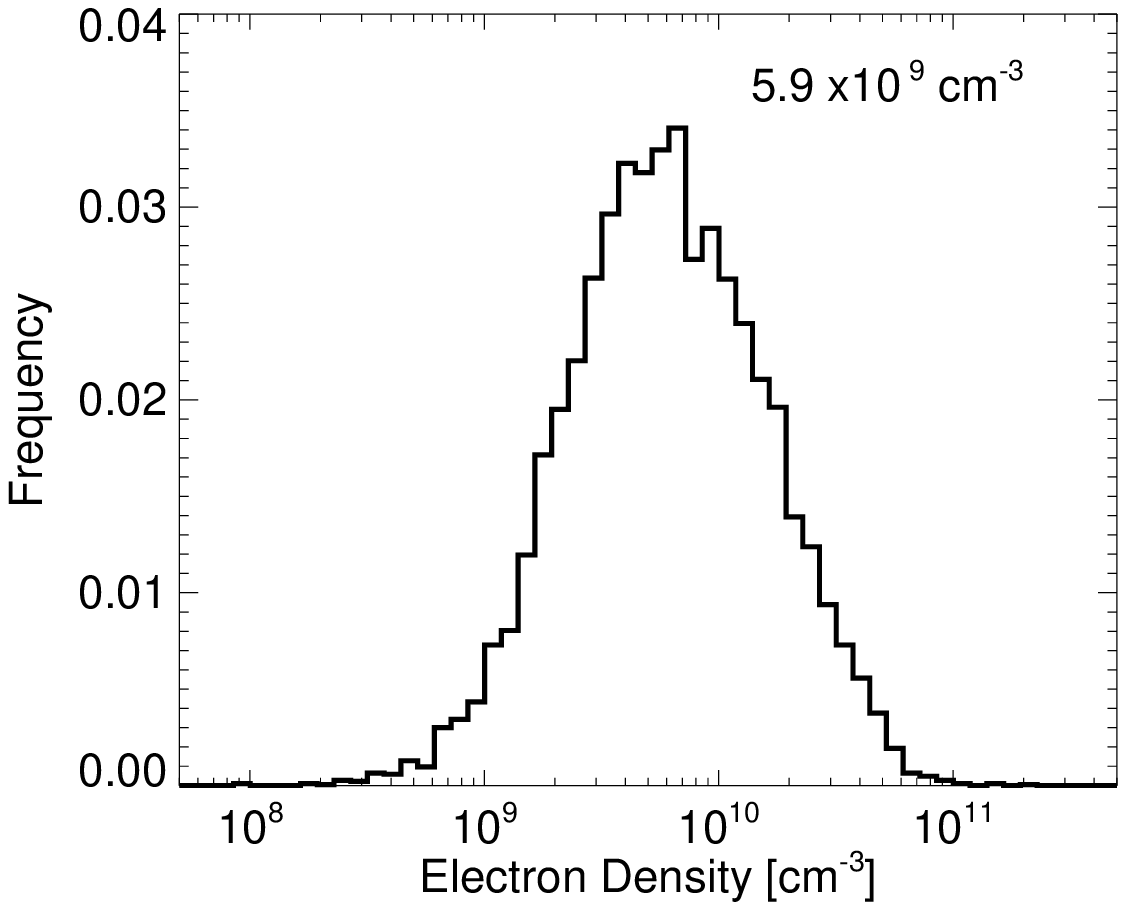}\\
\caption{Histograms of the density of the 4-8~keV loops, found from the
emission measure and thermal volume for 9,161 {\it RHESSI}
microflares.}\label{fig:histne}
\end{figure}

The histograms of the fitted temperature and emission measure for the
events with good background subtractions and thermal fits, 9,161
microflares, are shown in the top row of Figure~\ref{fig:histtemgeb}. This
is about a third of the total sample but shows nearly all of the events with
good background subtraction, flare signal-to-background and visibility
forward fits (9,693 microflares). The majority of the temperatures found lie
within a tight range of about 10 to 15 MK, with the median temperature of
around 13MK. The emission measures vary considerably more than the
temperatures, having a range covering over two orders of magnitude
between $10^{45}$ to $10^{47}$~cm$^{-3}$. The median emission
measure is $3\times10^{46}$~cm$^{-3}$. Also shown in
Figure~\ref{fig:histtemgeb} are the average ratio of the error in the fit to
the fitted parameter. For the temperatures this statistical error in the fit is
$<1\%$ and is approximately constant for the temperatures found. The
error in the emission measure is $\approx10\%$ at $10^{45}$~cm$^{-3}$
but drops to $\approx1\%$ at $10^{47}$~cm$^{-3}$. The larger relative
error in the events with smallest emission measure is due to the emission
measure being directly proportional to the thermal emission model, and so
are events with small noisy spectrum. The range of these parameters is
discussed further in \S\ref{sec:cor}. With these emission measures and the
volumes of the emitting plasma (see \S\ref{sec:vis}) an estimate of the
electron density can be made. The histogram of these densities is shown
in Figure~\ref{fig:histne} and range from $6\times10^{8}$ to
$6\times10^{10}$~cm$^{-3}$ with median value of $6\times
10^{9}$~cm$^{-3}$. This is larger than typical coronal conditions but
reasonable for a flaring loop \citep{phillips1996,gallagher1999}.

\begin{figure}\centering
\plotone{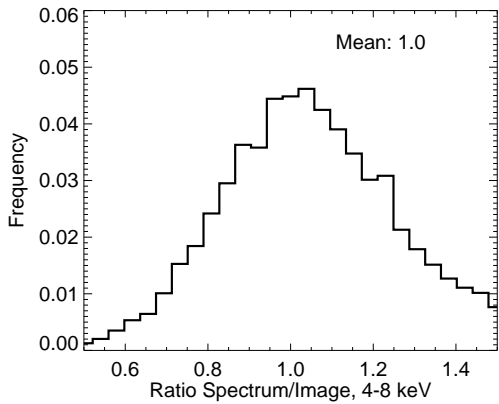}\\
\caption{Histogram of the ratio of the 4-8 ~keV flux found from the
fitted spectrum model and the image model loop.}\label{fig:histrat48}
\end{figure}

\begin{figure*}\centering
\includegraphics[scale=.8]{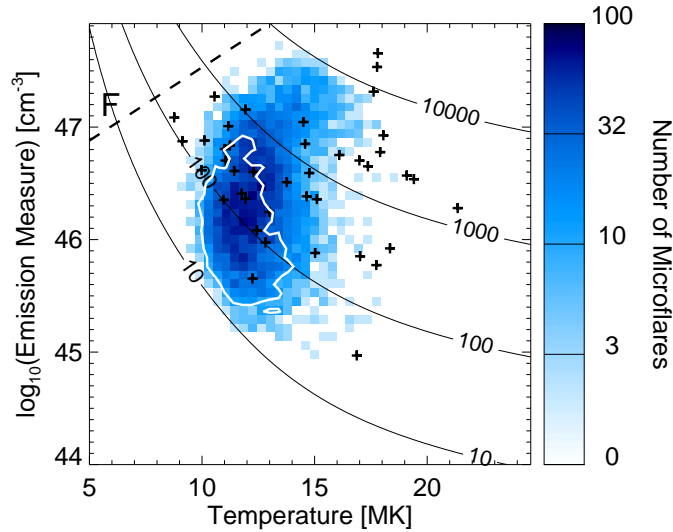}\\
\caption{Microflare temperature against emission measure. The
correlation plot is the microflares in this study with black crosses
representing the results of a previous {\it RHESSI} study
\citep{battaglia2005}. The dashed straight line represents the correlation
found by \citet{feldman1996big} from {\it GOES} and {\it Yohkoh/BCS}. The
numbered curved lines are of constant count rate per detector in 4-8~keV
for the thermal model, as a function of temperature and emission
measure. The white contour show  the events that have a {\it GOES}
temperature $<10$~MK in Figure \ref{fig:goesvsrhessi}.} \label{fig:emvst}
\end{figure*}

\begin{figure}\centering
\includegraphics[scale=0.6]{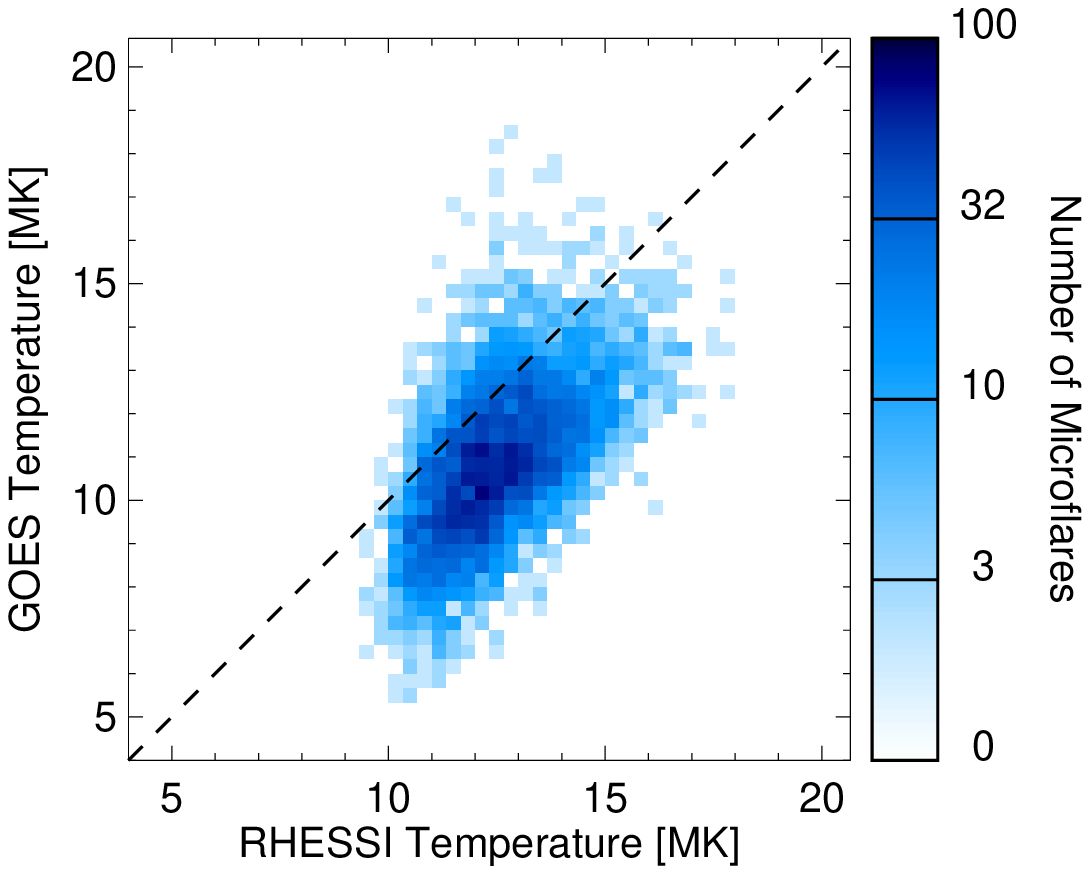}\\
\includegraphics[scale=0.6]{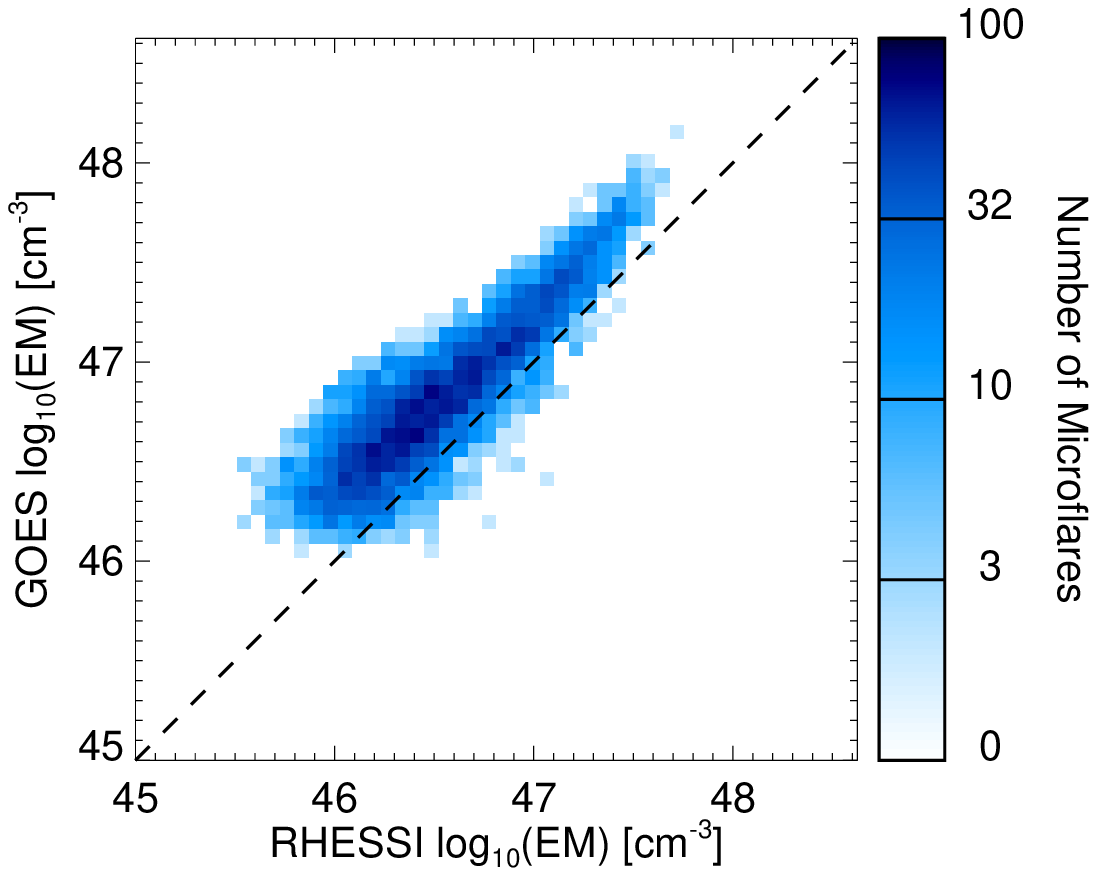}\\
\caption{Correlation plots of temperature (top) and emission measure
(bottom) from {\it RHESSI} and {\it GOES} for the same peak time
in {\it RHESSI} 6-12~keV for 6,740 microflares.
The dashed lines indicate a one-to-one correlation.}
\label{fig:goesvsrhessi}
\end{figure}

\begin{figure}\centering
\includegraphics[scale=0.6]{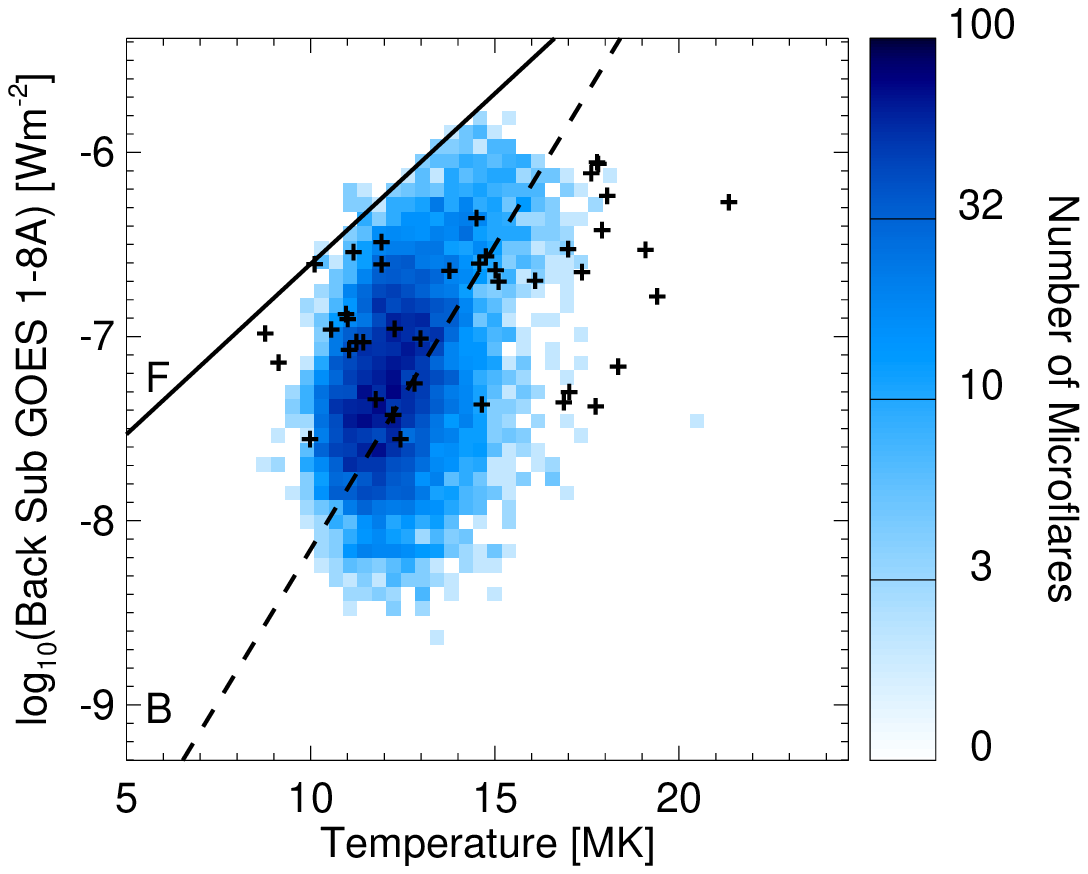}\\
\includegraphics[scale=0.6]{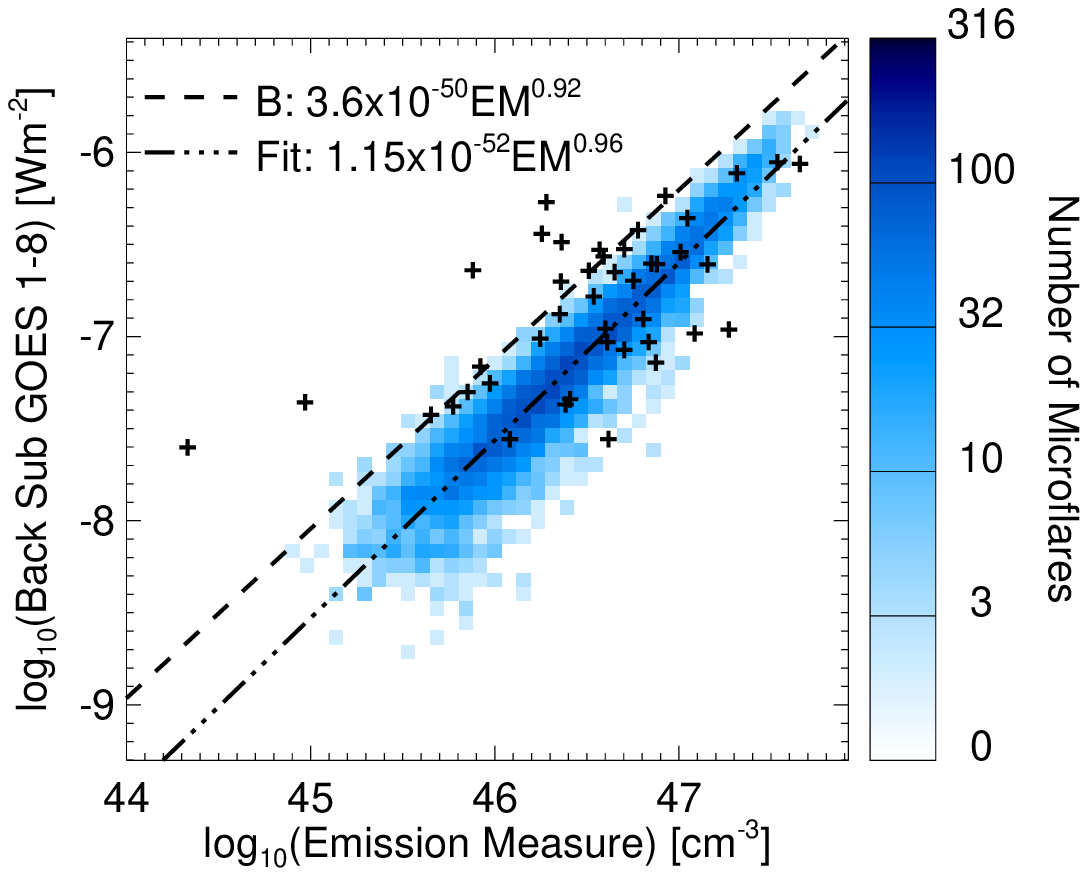}\\
\caption{Microflare temperature (top) and emission measure (bottom)
against the background subtracted {\it GOES} 1-8\AA~flux.
The correlation plots are the microflares in this study, with
the dot-dashed line (bottom panel)
the fit to this data. The black crosses and dashed
line were found from a previous {\it RHESSI} study \citep{battaglia2005}.
The solid line (top panel) represents the correlation found by
\citet{feldman1996big}.} \label{fig:goesvsemt}
\end{figure}

With these fitted thermal components we can estimate the 4-8~keV flux
from these spectrum fits and compare it to the flux derived from the
imaging in \S\ref{sec:vis}. This provides a consistency check to verify that
these two vastly different analysis techniques recover similar fluxes. The
histogram of the ratio of the flux found from the spectrum model to the
image value is shown in Figure~\ref{fig:histrat48}. The median of these is
$1.0$ with some spread about this value. This is expected as different
detectors were used for imaging and spectral analysis (additional use of
detector 5 in imaging) and the imaging calculation uses only the diagonal
elements of {\it RHESSI's} detector response matrix, \citep{smith2002},
whereas the spectrum fitting uses the full response matrix.

\subsection{Non-thermal Parameters}\label{sec:nontherm}

Histograms of the index $\gamma$ and break energy $\epsilon_\mr{B}$ of
the non-thermal broken power-law are shown in the bottom row of
Figure~\ref{fig:histtemgeb}. The power-law index $\gamma$ has values
mostly ranging over 4 to 10 with the median about 7. This is considerably
steeper than large flares observed by {\it RHESSI}, as discussed in previous
{\it RHESSI} microflare work \citep{krucker2002,benz2002}. The break
energy $\epsilon_\mr{B}$ ranges over 7~keV to 12~keV with the median
being about 9~keV, which is smaller than is found for larger flares
\citep{psh2005}. In larger flares, the lower energy non-thermal emission
would be masked by the thermal emission to tens of keV. The steep power
laws starting at low energies leads to a strong selection effect. This is
because such steep power-laws, extending down to energies where there
are spectral lines \citep{phillips2004}, are difficult to distinguish from a
thermal component. A conservative approach has been taken here to
remove any events where there is an ambiguity between the thermal and
non-thermal components, so we discount any events any events with
$\epsilon_\mr{B} \leq 7$~keV. The result is that we have only 4,236
microflares, about a fifth of the total sample.

Also shown in Figure~\ref{fig:histtemgeb} are the average ratios of the
error in the fit to the fitted parameter. For the power-law index $\gamma$
this statistical error in the fit is $\approx4\%$ for $\gamma<8$ and
increases to $\approx10\%$ for $\gamma>8$. This increase in the error
shows the greater uncertainty in trying to fit steep spectrum. The error in
the break energy is $\approx10\%$ at about $\epsilon_\mr{B}=7$ keV and
decreases to $>1\%$ by $\epsilon_\mr{B}=12$ keV. The large errors at low
break energies shows the greater uncertainty in trying to separate the
thermal and non-thermal components below 10 keV. These non-thermal
parameters are discussed further in \S\ref{sec:nonthermpow}.

\begin{figure}\centering
\includegraphics[scale=0.6]{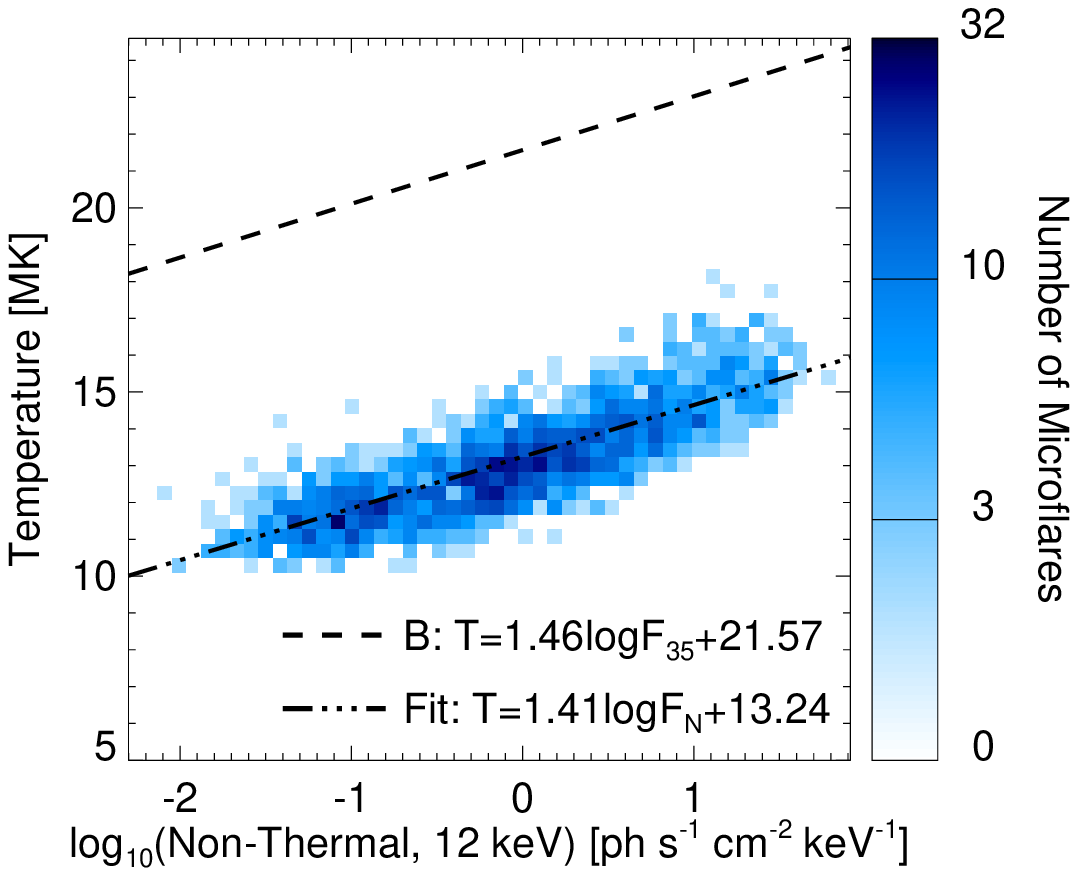}\\
\includegraphics[scale=0.6]{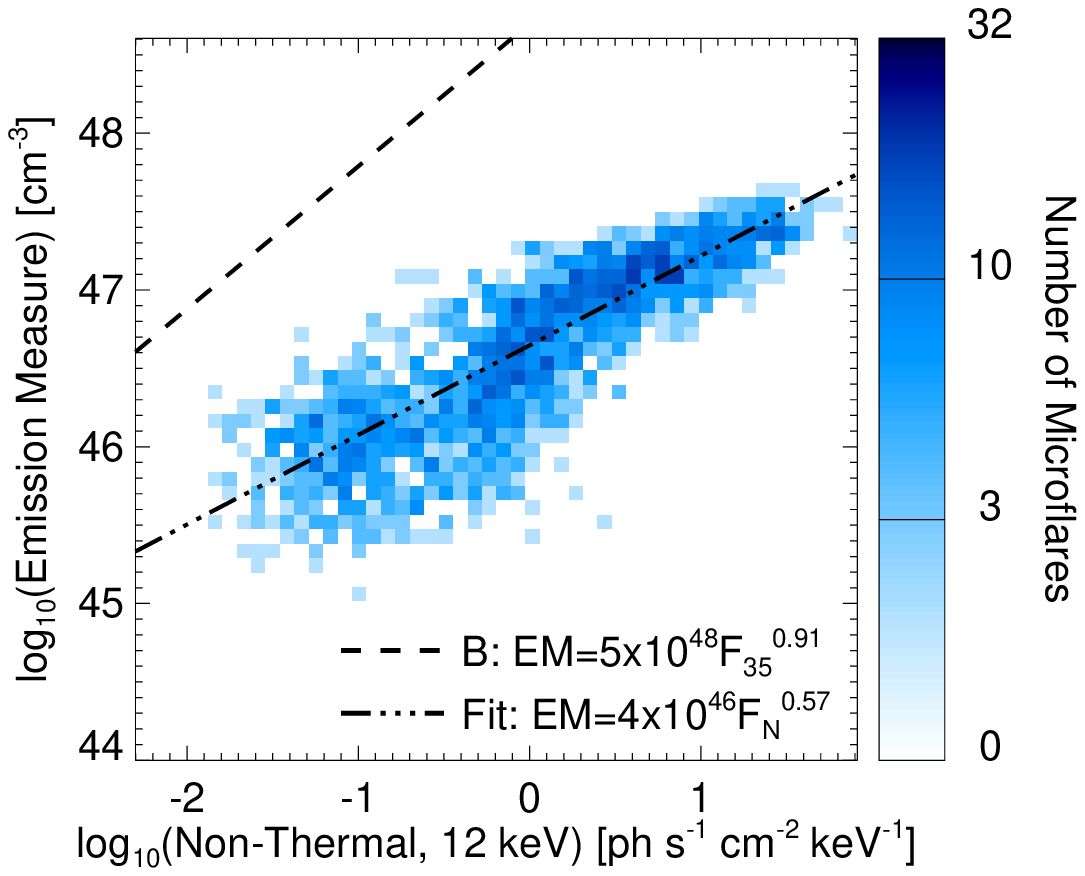}\\
\caption{{\it RHESSI} non-thermal photon flux at 12 keV, from the fitted
broken power-law, against fitted temperature (top) and emission
measure (bottom). The correlation plots are the microflares in this
study, the fit to this data is the dot-dashed line.
The dashed line is from a previous
{\it RHESSI} study using the non-thermal flux at 35~keV
\citep{battaglia2005}.} \label{fig:phnnvsemt}
\end{figure}

\subsection{Correlation Between Parameters}\label{sec:cor}

Figure~\ref{fig:emvst} plots the temperature against emission measure for
the {\it RHESSI} microflares.  There is no clear correlation between these
parameters. Also shown are numbered contours indicating constant count
rate per detector over 4-8~keV for the thermal model as a function of
temperature and emission measure. All of the microflares lie between the
10 and 10$^4$~counts s$^{-1}$ contours, consistent with the
non-background subtracted count rates for good fits shown in
Figure~\ref{fig:histrate48}. Although any temperature and emission
measure between these contours could be expected, the temperatures lie
in a tight range, mostly between 10~MK and 15~MK, with almost all
possible emission measures, from $10^{45}~\mr{cm}^{-3}$ to
$10^{47}~\mr{cm}^{-3}$, for this temperature range found. The model of
the thermal emission is directly proportional to $EM$ and increases with
larger $T$, though not directly, with the continuum rising and flattening
and the line features becoming more prominent \citep{tandberg1988}. This
results in the errors in the temperature and emission measure being
anti-correlated. The thermal model also includes spectral features and
they provide additional emission, particularly from the Fe K-shell over 4-8
~keV, for temperatures above $8$~MK \citep{phillips2004}. The fact that
only temperatures above $8$~MK have been found is more suggestive of a
selection effect primarily affecting the temperatures and not a physically
significant discovery of microflares with a lack of low temperatures and
high emission measures. This selection effect is consistent with {\it
RHESSI's} sensitivity being temperature-dependent. The combination of
this greater sensitivity to hotter plasma and the differential emission
measure DEM decreasing as the temperature increases could explain the
tight range of temperatures found with the peak of this sensitivity for {\it
RHESSI} shutter-out mode occurring in this temperature range. Note that
this selection effect essentially does not reject any event detectable by {\it
GOES}.

Microflares analyzed in a previous {\it RHESSI} spectral study of flares of
all scales, 42 microflares out of a sample of 85 flares,
\citep{battaglia2005} are shown as the crosses in Figure~\ref{fig:emvst}
and also showed no correlation. These events specially chosen to cover a
wide range of {\it RHESSI} flare magnitudes were analyzed during the peak
in 12-25~keV. This is earlier in the flare phase than for our survey. Thus
they have correspondingly higher temperatures and lower emission
measures. Another recent study of 18 microflares from a single active
region \citep{stoiser2007} found similar results to the \citet{battaglia2005}
study.

The dashed line in Figure~\ref{fig:emvst} is the correlation found using
soft X-ray observations with {\it GOES} and {\it Yohkoh/BCS}
\citep{feldman1996big}, for all size of flares from A through to X class, not
just microflares. This correlation has a order of magnitude spread in
emission measure and only becomes apparent when a large range of flare
magnitudes are included; for only small A-class events the correlation was
in the opposite direction \citep{feldman1996small} with the emission
measure decreasing with increasing temperature. In comparison the {\it
RHESSI} data, which is observed at higher energies than
\citet{feldman1996big}, show higher temperature and/or lower emission
measures. This is consistent with the DEM peaking at temperatures lower
than those observed with {\it RHESSI} and closer to those lower
temperatures observed by {\it GOES} and {\it Yohkoh/BCS}. This was found
to be case for the DEM of a large flare observed in soft X-rays with {\it
GOES} and {\it Yohkoh/SXT} and in hard X-rays with {\it Yohkoh/HXT}
\citep{aschalex2001}. For {\it RHESSI} emission measures above
$8\times10^{46}~\mr{cm}^{-3}$ there is the hint of a similar correlation
with temperature but it is obscured by the temperature selection effect at
lower emission measures. Further studies of solar flare temperature and
emission measure suggested that the emission measure of EUV nanoflares
approximately scales as $T^5$ \citep{asch2000}, but using various studies
over larger scales suggested the emission measure may scale as $T^4$
\citep{asch2007}. These scalings are only approximate as there is large
scatter about the correlation line. Looking over larger ranges with different
instruments helps to reduce the overall influence of the selection effects
but there is still ambiguity as to how these parameters scale.

To gain a better understanding of the {\it RHESSI} temperature and
emission measure, we have also calculated the {\it GOES} temperatures
and emission measures for each of these microflares. This was done using
the same pre-flare time for background and peak time of emission in
6-12~keV as was used in the {\it RHESSI} analysis. This was possible for
6,740 microflares, removing those events for which a {\it GOES}
temperature and emission measure were not reliably calculable. The
resulting correlation plots are shown in Figure~\ref{fig:goesvsrhessi}. The
bottom right hand of this correlation plot (the lowest {\it GOES}
temperature for each {\it RHESSI} temperatures) does seem to be directly
proportional, except the {\it RHESSI} temperatures are about 5~MK higher.
The rest of the plot is again dominated by the selection effect in the {\it
RHESSI} temperatures. The events with {\it GOES} temperature $<10$~MK
have been highlighted by the white contour in Figure \ref{fig:emvst}. If the
assumption is that the {\it RHESSI} temperate estimate is too high in these
events, and the emission measures is consequently too small, then this set
of highlighted events should be moved upwards and to the left in Figure
\ref{fig:emvst}. This shift produces a clearer hint of the previously found
correlation between temperature and emission measure.

\begin{figure}\centering
\includegraphics[scale=0.7]{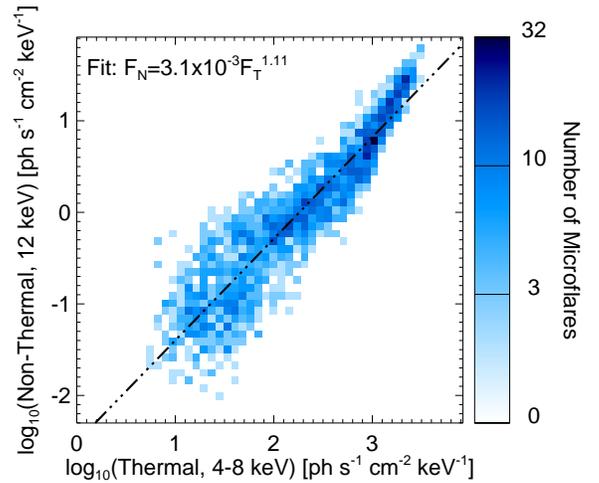}\\
\caption{
{\it RHESSI} non-thermal flux at 12 keV, from the fitted broken power
law, against the the thermal flux over 4-8~keV, from the thermal model. The
dot-dashed line is the fitted correlation line.} \label{fig:thvsnn}
\end{figure}

The emission measures in Figure~\ref{fig:goesvsrhessi} do not show any
such temperature selection effect and are nearly directly proportional, with
the {\it GOES} emission about twice that observed in {\it RHESSI}. Again
this will be due to {\it RHESSI} observing in a temperature range which is
higher than the peak temperature in the DEM. The proportionality between
the {\it RHESSI} and {\it GOES} emission measures on face value suggests
a similarly steep DEM in all the microflares but this might be arising from
the relative instrumental sensitives. It is important to remove these
instrumental effects to recover the underlying DEM but this is a
complicated process and has only been successful for individual large
flares, \citep[c.f.][]{aschalex2001}.

Other studies \citep{feldman1996big,battaglia2005} have also
investigated how the temperature and emission measure relate to the {\it
GOES} flux of the events. The {\it RHESSI} microflare temperature and
emission measure plotted against each event's corresponding
background-subtracted {\it GOES}~1-8\AA~is shown in
Figure~\ref{fig:goesvsemt}. Again there is a hint of a correlation between
the {\it RHESSI} temperature and {\it GOES} flux for the events above
B-class ($10^{-7}~\mr{Wm}^{-1}$), but it is obscured in smaller events
again due to the temperature selection effect. The previous study of {\it
RHESSI} flares \citep{battaglia2005} found a correlation when using large
flares in addition to microflares. This correlation and the microflares in
their sample are shown in Figure~\ref{fig:goesvsemt}, by the dashed line
and crosses, and is steeper than found using {\it GOES} and {\it
Yohkoh/BCS} \citep{feldman1996big}. The hint of a correlation in our
microflare study scales in a manner closer to \citet{feldman1996big} than
\citet{battaglia2005}, although the temperatures found are consistently
higher.

There is a clear correlation between the emission measure and {\it GOES}
flux, which can be fitted as $F_\mr{G}=1.15\times10^{-52}EM^{0.96}$,
with $F_\mr{G}$ in Wm$^{-2}$ and $EM$ in cm$^{-3}$. A similar result
was also found in the \citet{battaglia2005} study, fitted over a large flare
to microflare range finding $F_\mr{G}=3.6\times10^{-50}EM^{0.92}$. The
consistently lower emission measures in this study are again due to this
analysis occurring earlier in the microflare than in our study.

We can also investigate how the non-thermal emission relates to the
thermal parameters, which is shown in Figure~\ref{fig:phnnvsemt}. Here
the non-thermal flux $F_\mr{N}$, in units of photon flux at Earth
$\mr{s}^{-1} \mr{cm}^{-2} \mr{keV}^{-1}$, is taken from the fitted broken
power-law parameters at 12~keV. This energy is used as it is above the
break energy $\epsilon_\mr{B}$ in all events but is still at an energy for
which we observe non-thermal emission. Both the temperature and
emission measure scale with the non-thermal emission. The correlation
with the temperature is fitted as $T=1.41\log{F_\mr{N}}+13.2$, with $T$ in
MK. A similar correlation was found in the \citet{battaglia2005} study
using the flux at 35~keV for the non-thermal flux, since larger flares were
also analyzed, finding $T=1.46\log{F_\mr{35}}+21.57$. We find that the
scaling of the emission measure is flatter,
$EM=4\times10^{46}F_\mr{N}^{0.57}$, compared to the previous study
$EM=5\times10^{48}F_\mr{35}^{0.91}$ \citep{battaglia2005}, again $EM$
in units of cm$^{-3}$.

In Figure~\ref{fig:thvsnn} this non-thermal photon flux at 12~keV
$F_\mr{N}$ is plotted against the model thermal flux over 4-8~keV,
$F_\mr{T}$. The thermal and non-thermal emission correlate, with the fit
being $F_\mr{N}=3.4\times10^{-3}F_\mr{T}^{1.09}$, but with a greater
scatter in the smaller events. At higher fluxes the correlation steepens,
suggesting that there might a greater proportion of non-thermal emission
relative to thermal emission in larger flares. However, this may just be an
instrumental effect, arising from detector pileup and livetime issues in
these larger events prior to the shutters deploying.

\section{Energy Distributions}\label{sec:eng}

\begin{figure*}\centering
\includegraphics[scale=0.9]{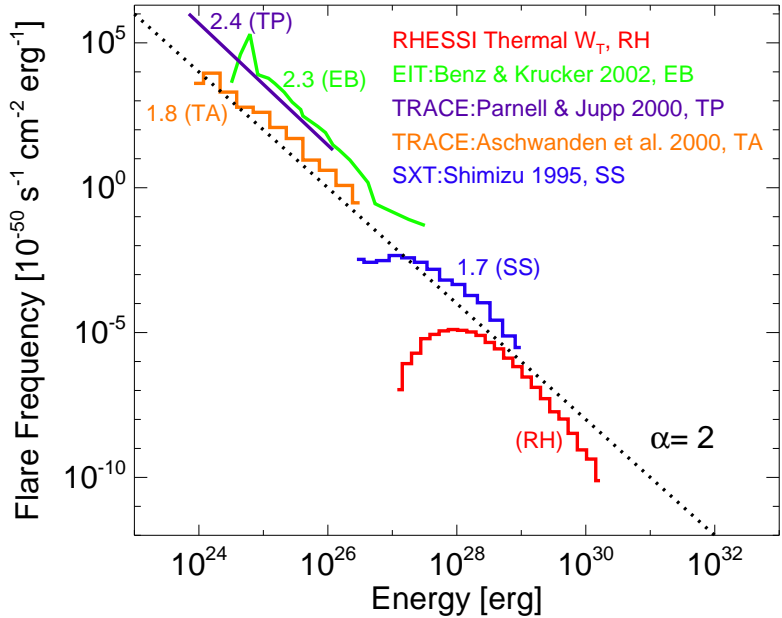}\\
\caption{Frequency distribution of the thermal energy,  of 9,161 {\it
RHESSI} microflares (RH) in context of thermal energy distributions of
nanoflares (TA, TP and EB) and active region transient brightenings (SS).
The dotted line indicated a power-law index of $\alpha=2$.}
\label{fig:thermdist} \end{figure*}

\begin{figure}\centering
\plotone{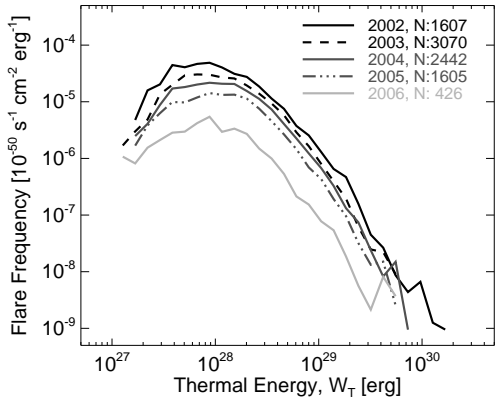}\\
\caption{Frequency distribution of the microflare thermal energy, each line
indicating a different year. The number quoted next to the year is
the number of events from each year used in the distribution.}
\label{fig:thermdistpy}
\end{figure}

\subsection{Thermal Energy Frequency Distributions}

Using the volumes found in \S\ref{sec:vis} and temperature and emission
measure found in \S\ref{sec:ospex} the thermal energy $W_\mr{T}$, over
the time of peak emission in 6-12~keV, can be calculated via equation
(\ref{eq:therm}). The energies range from $10^{26}$~erg to $10^{30}$~erg
with the median being about $10^{28}$~erg. The frequency distribution
(number of events per energy bin range, area of solar disk and duration of
observation period) of this energy for 9,161 microflares is shown in
Figure~\ref{fig:thermdist}. The resulting {\it RHESSI} thermal energy
distribution is not a clear power-law; it has a turn-over at low energies and
steepens at higher energies. These features are instrumental effects due
to missing the smallest events, with insufficient counts to either find the
events or successfully analyze them, and the largest events, due to
detector livetime issues before the attenuating shutters come in.
Considerably more small events are missing than large, since the
discrepancy from a power-law is greater at lower energies than high. As
this distribution deviates from a power-law it will not be fitted to obtain
the power-law index $\alpha$. Instead, Figure \ref{fig:thermdist}, shows
an $\alpha=2$ line, indicating the parts of the {\it RHESSI} thermal
distribution can be steeper and flatter, or larger and smaller, than
$\alpha=2$.

Compared to the previous distributions found for EUV nanoflares
\citep{krucker1998,asch2000,parnell2000,bk2002} and soft X-ray
active-region transient brightenings \citep{shimizu1995}, the {\it RHESSI}
energy distribution appears as an extension to these at higher energies.
This is both remarkable and deceptive since these distributions were found
for very different types of events, using various instruments and for
different periods during the solar cycle. For instance the {\it SXT} energies
\citep{shimizu1995} are from 291 brightenings in one active region over 5
days in August 1992, whereas the {\it EIT} \citep{krucker1998,bk2002} and
{\it TRACE} EUV quiet sun observation \citep{asch2000,parnell2000} were
found over about an hour each on 12 July 1996, 17 February 1998 and 16
June 1998 respectively. So there are two key issues that have to be taken
into account when looking at the energy distributions in
Figure~\ref{fig:thermdist}. First, various instruments were used, so
different components of the thermal energy will be observed and with
distinctive instrumental selection effects will influence each distribution.
This makes it difficult to determine whether these are similar events or
completely distinctive physical processes. Second, the distributions cover
different phases of the solar cycle. The previous studies show a snapshot
of the energy distribution of different small energy release features in the
solar corona whereas the {\it RHESSI} microflare energy distribution
represents 5 years of observations of the declining phase of the solar
cycle. The difficulty here lies in determining whether these snapshot
surveys demonstrate typical or unusual behavior. Using the {\it RHESSI}
thermal energies we can investigate how the energy distribution changes
with time by plotting the distributions for each year separately, as shown
in Figure~\ref{fig:thermdistpy}. These distributions have similar shapes
except that the normalization decreases by over an order of magnitude
between 2002 and 2006. So the non-{\it RHESSI} distributions in
Figure~\ref{fig:thermdist} could be shifted vertically by a considerable
amount if they were found during a different part of the solar cycle. This
invariance in the shape of the distribution during the solar cycle has been
found previously in both soft \citep{feldman1997,veronig2002} and hard
\citep{crosby1993,lu1993} X-rays, and also illustrates that the selection
effects on the RHESSI data do not vary with time.

\begin{figure*}\centering \includegraphics[scale=0.45]{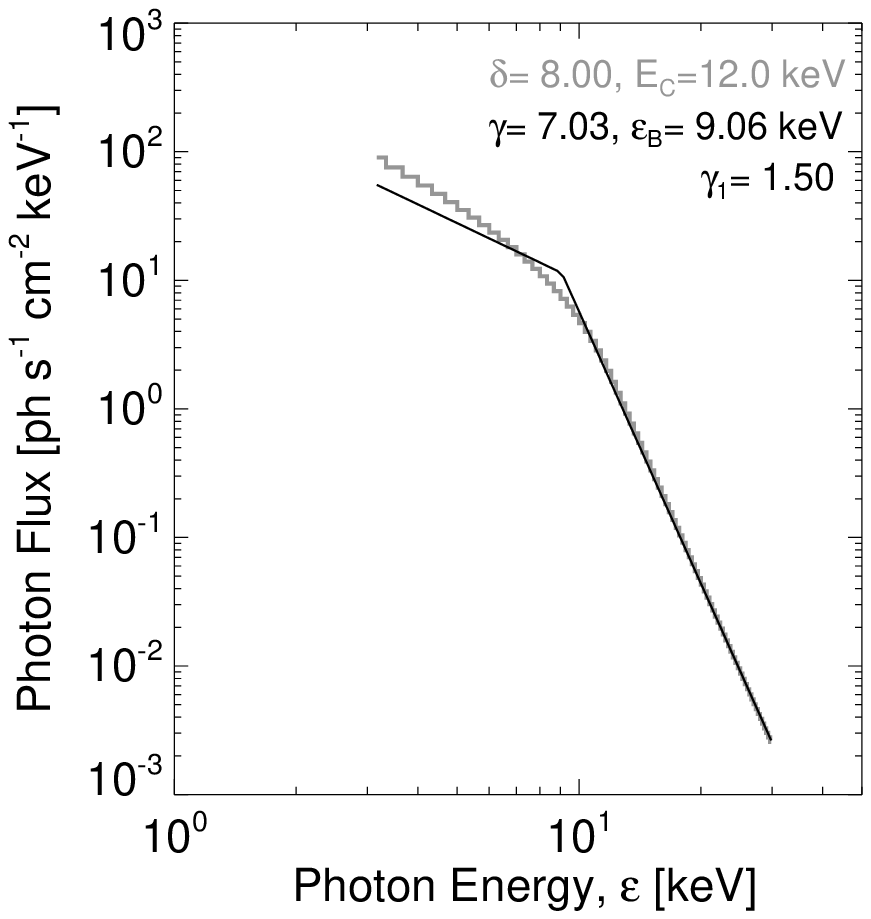}
\includegraphics[scale=0.45]{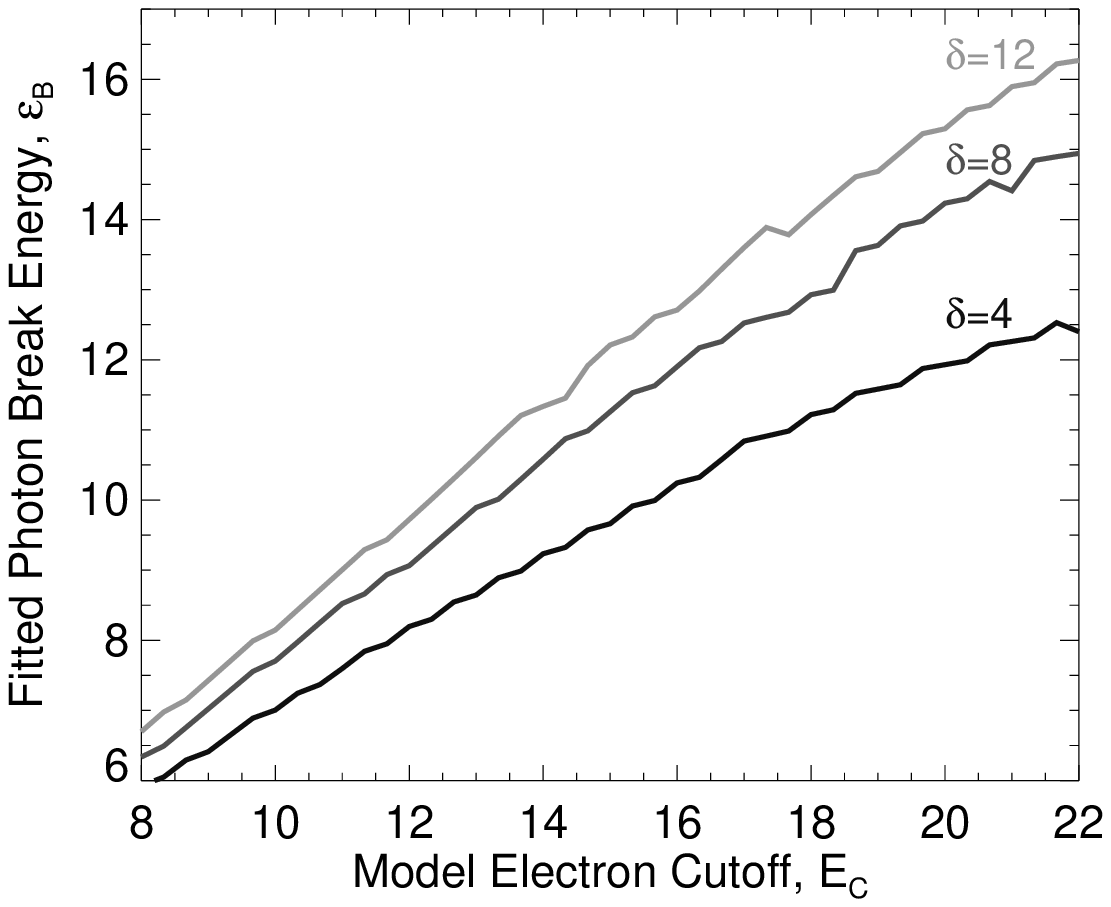}
\includegraphics[scale=0.45]{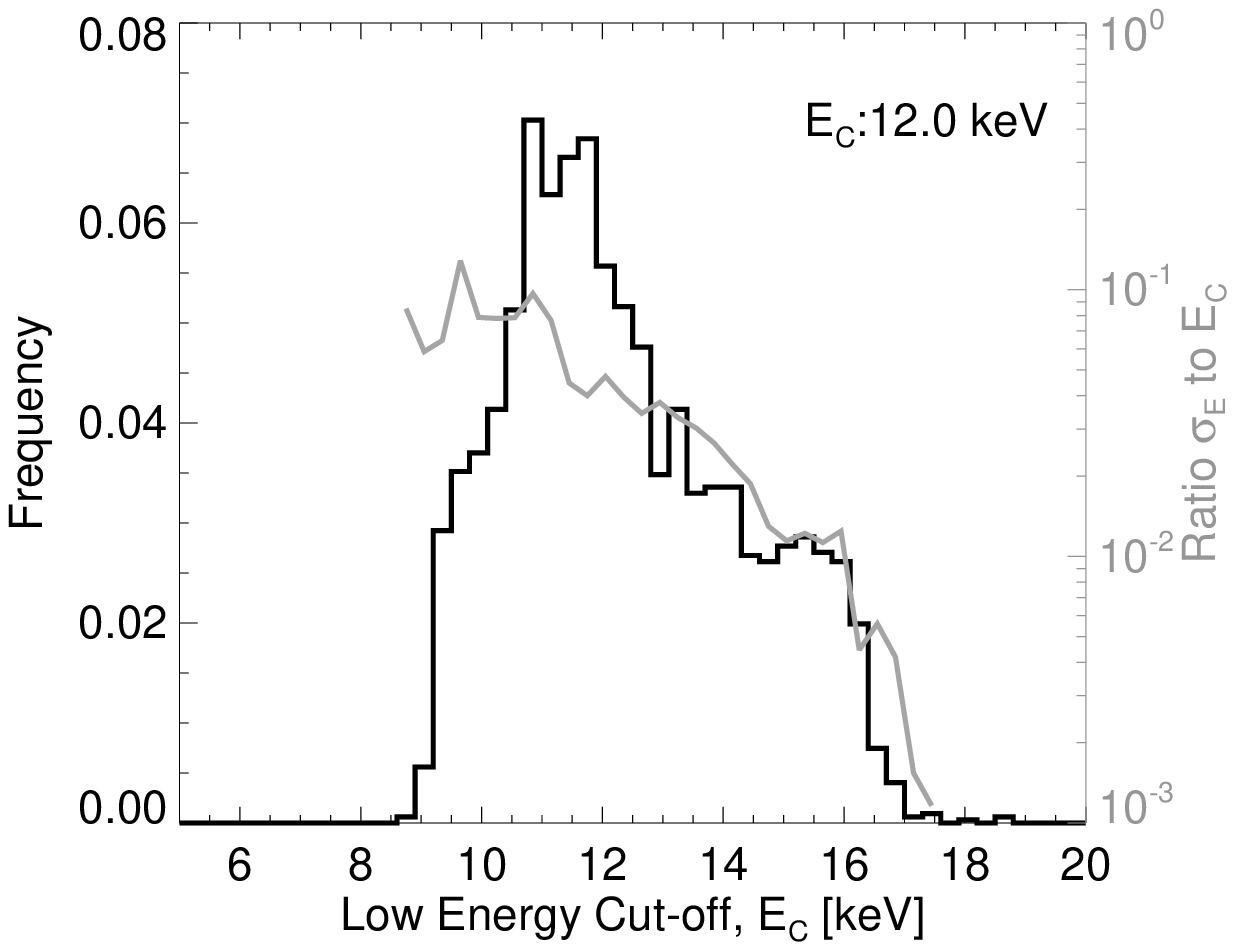}\\
\caption{(\emph{Left}) Broken power-law fit to photon spectrum from a
thick target model of a power-law electron distribution, using the code of
\citet{holman2003}. (\emph{Middle}) Fitted break energy
$\epsilon_\mr{B}$ of photon broken power-law against low energy cut-off
of electron distribution, the different lines indicate the different indices
$\delta$ of the electron distributions. (\emph{Right}) Histogram of
$E_\mr{C}$ for the 4,236 microflares shown in Figure \ref{fig:histtemgeb}.
The grey line refers to the righthand axis and shows the average ratio of
the error to $E_\mr{C}$ as a function of $E_\mr{C}$.}
\label{fig:thick2bpow} \end{figure*}

\subsection{Non-Thermal Power Frequency
Distributions}\label{sec:nonthermpow}

To calculate the power in accelerated electrons from the information about
the power-law in the photon spectrum, we use $\gamma$ and $I_\mr{0}$
(the power-law index and normalization), found in \S\ref{sec:ospex}, in
equation (\ref{eq:pow}). However there is ambiguity as to the low-energy
cut-off $E_\mr{C}$ because the observed photon spectrum depends only
weakly on it. In addition the {\it RHESSI} microflare spectrum covers both
the thermal and non-thermal energy ranges and so there is inherent
ambiguity. Previous studies used their X-ray threshold for $E_\mr{C}$
resulting in large uncertainties in the energy estimates
\citep{crosby1993,lin2001}. As {\it RHESSI} makes spectral measurements
down to the thermal component we can provide a better estimate of the
non-thermal energy content. In Part I of these papers a fixed $E_\mr{C}$
was used to provide a rough estimate of the non-thermal power at peak
time \citep{mfpart1}. The full spectrum fitting in \S\ref{sec:ospex} provides
a better estimate of the parameters required to calculate the non-thermal
power via equation (\ref{eq:pow}). However, despite this improvement we
only have an estimate of where the photon spectrum begins to flatten
$(\epsilon_\mr{B})$ and not the actual cut-off in the electron distribution
$E_\mr{C}$.

To obtain an estimate of $E_\mr{C}$ we have therefore investigated how
the fitted broken power-law model relates to the expected photon
spectrum from the electron distribution using the numerical integration
code of \citet{holman2003}. An example of this is shown in the left panel
of Figure~\ref{fig:thick2bpow}, where the photon spectrum for an electron
distribution with $\delta=8$ and $E_\mr{C}$ was calculated and then
fitted with a broken power-law in the same manner as was used for the
microflare spectrum. This broken power-law has $\gamma=7.03$,
expected as $\delta \equiv \gamma+1$ \citep{brown1971}, above a break
of $\epsilon_\mr{B}=6.06$~keV. Repeating this process for various values
of the electron distribution parameters $\delta$ and $E_\mr{C}$, shows
how the resulting fitted $\epsilon_\mr{B}$ scales with these parameters.
This is shown in the middle panel in Figure~\ref{fig:thick2bpow}, where
$\epsilon_\mr{B}$ is given against $E_\mr{C}$ for three different values of
$\delta$. For a single value of $\delta$, $E_\mr{C}$ approximately scales
linearly with $\epsilon_\mr{B}$ and this scaling steepens with increasing
$\delta$. So we can approximate this relationship to first order by linearly
fitting the relationship between $\epsilon_\mr{B}$ and $E_\mr{C}$ for
various $\delta$ and then linearly fitting how the parameters of the first fit
vary with $\delta$. This empirical relationship between the observed
parameters of the photon power-law $\gamma,\epsilon_\mr{B}$ and the
low energy cut-off of the electron distribution $E_\mr{C}$ can be found:
\begin{equation}\label{eq:ec} E_\mr{C}\approx
0.15\gamma+(1.86-0.04\gamma)\epsilon_\mr{B}-3.39. \end{equation}
\noindent Using equation (\ref{eq:ec}) the values of $\gamma$ and
$\epsilon_\mr{B}$ for the 4,236 microflares with trustworthy non-thermal
spectral fits result in a histogram of $E_\mr{C}$ for these events, shown in
the right panel in Figure~\ref{fig:thick2bpow}. The low energy cut-offs
range from 9 to 16~keV with the median being about 12~keV, with
generally $\epsilon_\mr{B} \approx 0.75 E_\mr{C}$. The uncertainty in
$E_\mr{C}$ can be seen in the average ratio of the error (found from the fit
errors in $\gamma$ and $\epsilon_\mr{B}$) to $E_\mr{C}$ also shown in
the right panel of Figure~\ref{fig:thick2bpow}. For $E_\mr{C}<12$ keV,
where the major of the event lie, the uncertainty is about $10\%$ and
drops to about $1\%$ for the few events with larger $E_\mr{C}$.

\begin{figure*}\centering \plottwo{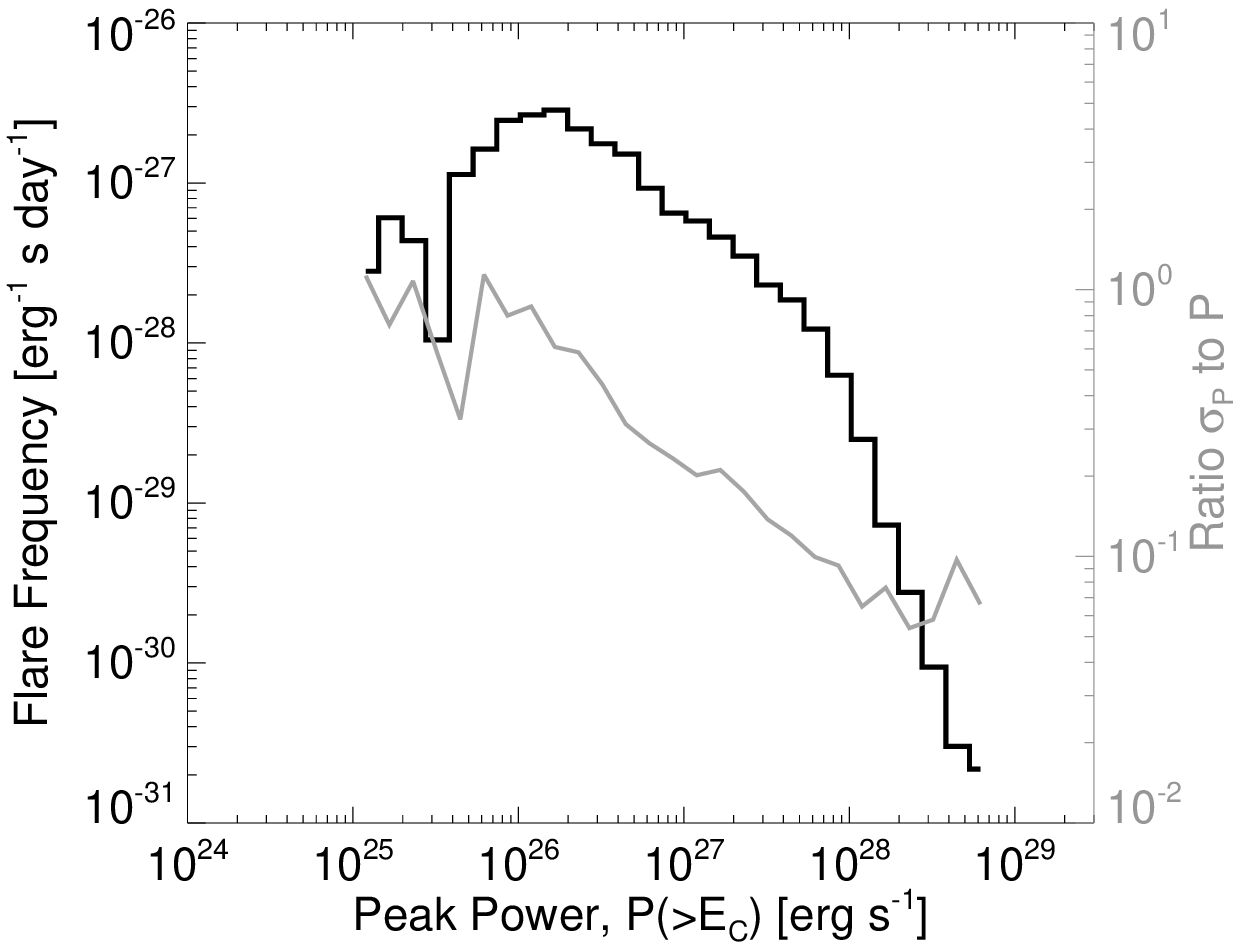}{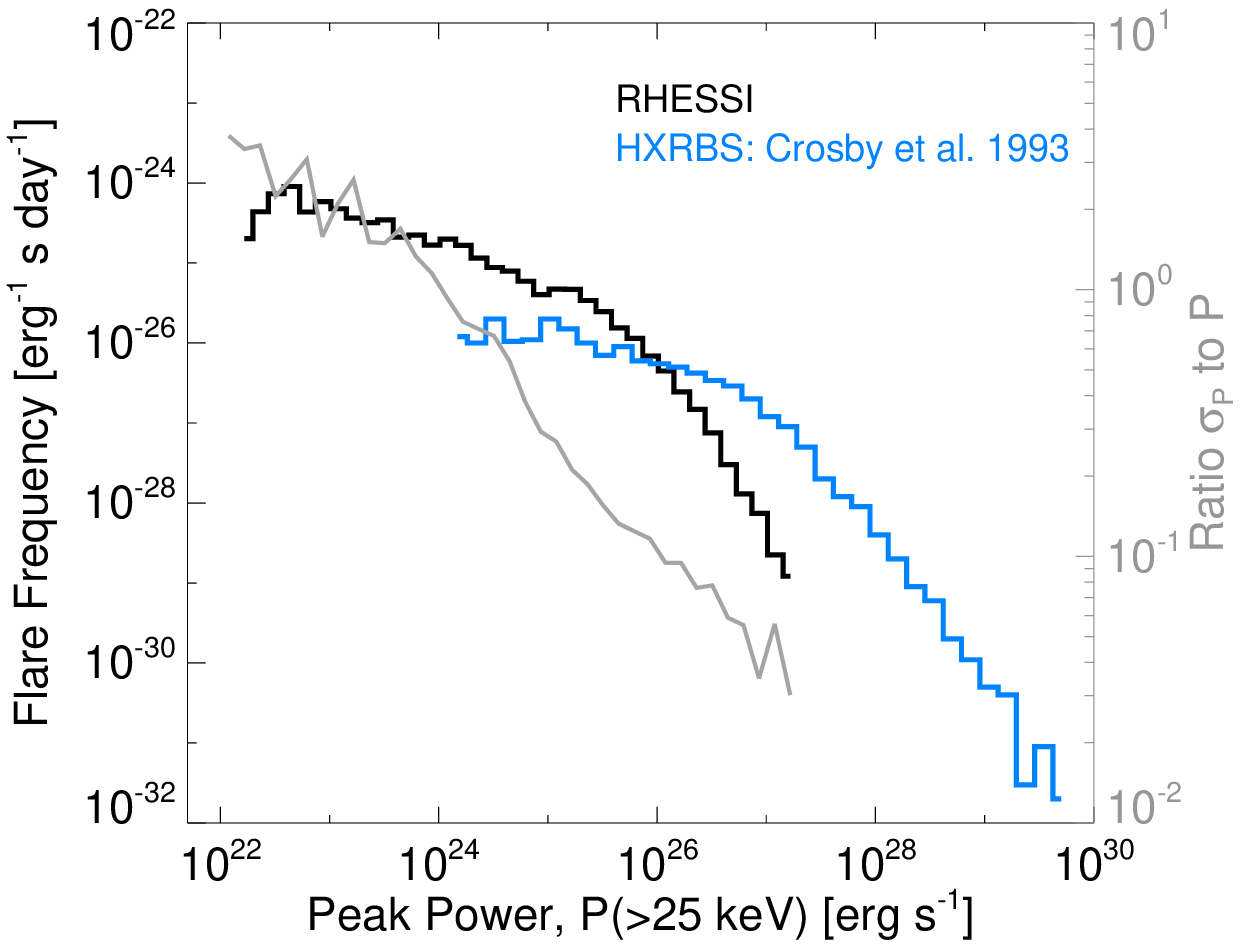}\\
\caption{Frequency distribution of the non-thermal power  of 4,236 {\it RHESSI}
microflares for $P(>E_\mr{C})$ (\emph{left}) and $P(>25\mr{keV})$, in
comparison to the distribution found by \citep{crosby1993} for large flares
(\emph{right}). The grey lines refers to the righthand axes and show the
average ratio of the error in the power  to the power, as a function of the
power.} \label{fig:nnthdist} \end{figure*}

These values of $E_\mr{C}$ can then be used in equation (\ref{eq:pow}) to
calculate the power in non-thermal electrons above $E_\mr{C}$, $P(\ge
E_\mr{C})$, the frequency distribution of which is shown in the left panel
in Figure \ref{fig:nnthdist}. Here the power ranges over $10^{25}$ to
$10^{28}$~erg~s$^{-1}$ with the median being about
$10^{26}$~erg~s$^{-1}$. Also shown is the ratio of the error to the power,
as a function of the power. The few microflares which show a power of
around $10^{28}$~erg~s$^{-1}$ have the smallest errors with
uncertainties about $10\%$. These powers seem relatively high for small
flares, as the {\it RHESSI} power estimates in large flares are $10^{27}$ to
$10^{30}$~erg~s$^{-1}$
\citep{holman2003,emslie2004,psh2005,sui2005,sui2007}, though it
should be noted that the non-thermal emission in microflares lasts for
$\approx 10$ seconds whereas it can last for tens of minutes in large
flares. The total non-thermal energy content in large flares is many orders
of magnitude larger than in microflares. The majority of the microflares
shown in Figure \ref{fig:nnthdist} show non-thermal powers considerably
smaller than this level, although with increasing uncertainty. The median
power of $10^{26}$~erg~s$^{-1}$ has about a $50\%$ error and the
smallest events at $10^{25}$~erg~s$^{-1}$ have almost $100\%$ error. It
is this large uncertainty in the power that results in this distribution
deviating from a power-law. Although the selection effects and biases will
affect this non-thermal distribution as seen in the thermal distribution,
these effects are hidden by the large uncertainties. Since it deviates from
a power-law, it has not been fitted using this model.

To allow comparison to the power distribution in electrons above 25~keV
$P(\ge 25)$  for 2,878 large flares found over 1980 to 1982 using {\it
SMM/HXRBS} \citep{crosby1993} the same power is estimated in the {\it
RHESSI} microflares, shown in the right panel in Figure \ref{fig:nnthdist}.
These powers using $E_\mr{C}=25$ keV for the {\it RHESSI} microflare
ranges over $10^{22}$ to $10^{27}$~erg~s$^{-1}$ with the median power
being about $10^{24}$~erg~s$^{-1}$. On  this basis the {\it RHESSI}
microflares cover events down to two orders of magnitude smaller than
those in the {\it SMM/HXRBS} study.  Again the error is smallest in the
largest power estimates. This $P(\ge 25)$ {\it RHESSI} distribution deviates
less from a power-law than the $P(\ge E_\mr{C})$ distribution. This is
because there is less uncertainty in the power estimate in the larger
events as the fixed $E_\mr{C}=25$ keV has no associated error.

\section{Discussion \& Conclusions}\label{sec:discons}

We have analyzed 25,705 microflares, successfully recovering trustworthy
spatial information about the thermal emission for 18,656 of them, as well
as the thermal spectral fit in 9,161 and the thermal and non-thermal fit in
4,236. The median values and ranges of each of the microflare parameters
are summarized in Table~\ref{table}. As found in Part I \citep{mfpart1} the
microflares occur only in active regions and this immediately suggests that
these events are ill-suited to heat the overall corona. It also helps to show
that the X-ray emission outside active regions, especially during periods of
quiet Sun, is considerably smaller than the smallest active-region flares
observed with {\it RHESSI}. This is confirmed by the {\it RHESSI} quiet Sun
study \citep{hannah2007} which found a limit on the flux over two orders
of magnitude smaller than that of the smallest microflares found here.

Forwarding-fitting the visibilities of {\it RHESSI} data is a fast and efficient
way to recover the spatial information \citep{hurfordvis}. We find that the
thermal emission (4-8~keV) of the microflares shows predominantly
loop-like structures that have a median width of 11$''$ and 32$''$ length.
Those events where no spatial information was recoverable were
predominantly those with the fewest counts, although some larger events
produce poor spatial information as there were other instrumental issues.
These spatial scales do not correlate with the magnitude of the microflare,
the 4-8 keV flux from the loop nor the GOES class, and so small flares are
not necessarily spatially small. This may be due to the prevalence of a
typical loop scale size associated with active regions, and it is
independent of the amount of energy the flare has deposited into heating
the material that evaporates to fill these loops. In the largest flares this
energy might be deposited over a larger area, evaporating material into
more loops, hence the larger structures and arcades that are observed.
The widths of these loops may not be as well resolved as the lengths, with
the observed single {\it RHESSI} loop possibly being several narrow and
long loops beside each other, as observed in microflare with {\it RHESSI}
and {\it Hinode/XRT} \citep{hannahaa2008}. The volume of these loops
can be estimated, the median value being $1\times10^{27}~\mr{cm}^{3}$.
This volume estimate assumes a cylindrical geometry and filling factor of
unity, meaning that our volumes, and hence thermal energies, are upper
limits, as this filling factor can be $<1$ \citep{cargill1997,takahashi2000}.

We find that microflares are hot, with a median temperature of 13~MK. In
these microflares the median emission measure is
$3\times10^{46}~\mr{cm}^{-3}$, which combined with the volume allows a
density of $6\times10^{9}~\mr{cm}^{-3}$ to be calculated. As the volume
may be an overestimate, this means that the density might be
underestimated. The presence of the spectral feature due to the Fe K-shell
transitions proves that there really is hot plasma present with $\mr{T}>
8$~MK \citep{phillips2004}. Correlations between the temperature and
other parameters are difficult to extract from our sample due to the
temperature selection effect, resulting in only a relative narrow range of
temperatures being sampled. Studying larger flares would aid by extending
the range of flare temperatures, but this would be complicated by
systematic effects related to {\it RHESSI}'s attenuating shutters. An
alternative would be to investigate this selection effect by comparing
model spectra for known temperatures and emission measures with those
found by fitting this spectrum once noise and the instrumental response
have been included. This has already been attempted for the {\it TRACE}
nanoflares study \citep{asch_parnell2002}, and it revealed a difference
between the observed parameters and the ``intrinsic'' ones. Another way
of investigating the instrumental response and biases is in comparison to
results from other instruments. We have found that the {\it GOES}
emission measures are typically a factor of 2 larger than those found with
{\it RHESSI}. This maybe due to the underlying DEM in microflares scaling
similarly but might also be due to the instrument's relative sensitivity. This
could be further studied by forward-fit modeling of the instrumental
responses to recover the underlying DEM in each event
\citep{aschalex2001}, instead of assuming a constant emission measure.

The thermal energies for the peak time in these microflares were found to
range from $10^{26}$~erg to $10^{30}$~erg with the median energy being
$10^{28}$~erg. The thermal energy distribution deviates from a power-law
at low and high energies but this can be explained by selection effects and
does not suggest that the underlying true distribution is not power-law.
The smallest events are missing as they do not have enough counts to be
found either above solar or instrumental background, successfully imaged
or spectrally analyzed. The larger events are missing from the distribution
due to {\it RHESSI}'s attenuating shutters, so either they are exclude from
our selection or have poor detector livetime. Although it is possible to
extend this energy distribution to higher energies by analyzing all large
flares observed with {\it RHESSI}, there will still be a strong instrumental
effect at the shutter transition due to detector livetime effects. The
statistical errors in the fitted parameters will affect the thermal energy
estimates but these have a relatively small effect compared to the
selection effects. The greatest uncertainty is the systematic error that
arises from the filling factor used to estimate the thermal volume: taking
an extreme filling factor of $f\approx10^{-4}$ \citep{cargill1997} would
have the effect of reducing the energies by a factor of 100. Therefore the
thermal energies quoted here about the time of peak emission in 6-12 keV
are upper limits. The comparison of the {\it RHESSI} thermal distribution to
other the thermal distributions of other transient coronal energy releases
is difficult as these events are from a limited sample of events over a very
short time period.

With {\it RHESSI} we are able to investigate the flattening of the
non-thermal photon spectrum and can empirically estimate the low energy
cut-off $E_\mr{C}$ in the electron distribution from the observed photon
spectrum. However we are only able to successfully fit the non-thermal
spectral component, in addition to the thermal component, in 4,236
events. The difficultly in successfully finding and fitting this non-thermal
emission arises from the steep spectra that start close to the thermal
spectral features about 7~keV. So, although the non-thermal component
was only successfully fitted in a minority of events, it does not mean that
there is a lack of accelerated electrons in the others. It does mean
however that the relatively large uncertainties in the non-thermal
parameters will result in large uncertainties in $E_\mr{C}$ and the
non-thermal power estimates. These large uncertainties and the biases
introduced by rejecting many events due to poor spectra fits can be easily
seen in the histograms of the non-thermal parameters $\gamma$ and
$\epsilon_\mr{B}$. We find that the microflares typically have non-thermal
emission represented by a broken power-law with index $\gamma=7$,
break at 9~keV. We can estimate $E_\mr{C}=12$~keV, for breaks in the
photon spectrum above 7~keV.

These non-thermal parameters were used to calculate the power in
accelerated electrons above $E_\mr{C}$, which range over $10^{25}$ to
$10^{28}$~erg~s$^{-1}$ with a median value of $10^{26}$~erg~s$^{-1}$.
This distribution again deviates from a power-law at low and high energies
due to the uncertainties in the power estimates and the rejected events
due to poor spectra fits rather than the selection effects present in the
thermal distribution. The uncertainties in the few events with the largest
non-thermal component are only $\sim10\%$ and so this cannot explain
the fact that in some of these small flares the rate of energy release is
comparable to larger flares, $10^{27}$ to $10^{30}$~erg~s$^{-1}$
\citep{holman2003,emslie2004,psh2005,sui2005,sui2007}. However the
non-thermal emission in microflares lasts for only $\approx 10$s seconds
whereas it can last for tens of minutes in large flares. The non-thermal
energy content in large flares is thus many orders of magnitude larger than
in microflares. To make a direct comparison of these {\it RHESSI}
microflare results to the peak non-thermal power distribution found for
large flares by \citet{crosby1993}, the same instrumental $E_\mr{C}=25$
keV has to be used. On this basis the {\it RHESSI} microflare peak powers
extend to two orders of magnitude smaller than the \citep{crosby1993}
study. The true non-thermal power in the \citet{crosby1993} flares cannot
be determined as the spectrum was not observed to sufficiently low
energies and so that study could have easily under-estimated the
non-thermal energy content. We conclude that the instantaneous
non-thermal power in a microflare can be surprisingly large.

\begin{deluxetable*}{lr|cc} \tablecolumns{4} \tablewidth{0pc}
\tablecaption{ \label{table}Median and range of {\it RHESSI} microflare
parameters.}
  \tablehead{\multicolumn{2}{l}{Microflare Parameter} &
  \colhead{Median Value} & \colhead{Range (5\% to 95\%)}}
   \startdata
   Duration\tablenotemark{a} & $D$ &5.4 mins & 2.2-15 mins\\
  Temperature & $T$ & $12.6$~MK &  $10.7-15.5$~MK\\
  Emission Measure & $EM$ & $3\times10^{46}$~cm$^{-3}$ &
  $4\times10^{45}-2\times10^{47}$~cm$^{-3}$ \\
  Thermal Loop Width & $w$ & 8~Mm ($11''$) &  3-20~Mm ($4-28''$)\\
  Thermal Loop Length & $l$ & 23~Mm ($32''$) &  7-77~Mm ($10-107''$)\\
  Thermal Loop Volume & $V$ & $1\times10^{27}$~cm$^{3}$ &
  $5\times10^{25}-2\times10^{28}$~cm$^{3}$ \\
 Density & $n_\mr{e}$ & $6\times10^{9}$~cm$^{-3}$ &
 $8\times10^{8}-3\times10^{10}$~cm$^{-3}$\\
  Power-law Index & $\gamma$ & 7 &  $4-10$\\
  Break Energy & $\epsilon_\mr{B}$ & 9~keV &  $7-12$~keV\\
  Low Energy Cut-off & $E_\mr{C}$ & 12~keV &  $9-16$~keV\\
  Thermal Energy\tablenotemark{b} & $W_\mr{T}$ & $10^{28}$~erg &
  $10^{26}-10^{30}$~erg\\
  Non-thermal Power\tablenotemark{b}& $P_\mr{N}(\geq E_\mr{C})$ &
  $10^{26}$~erg s$^{-1}$&  $10^{25}-10^{28}$~erg s$^{-1}$\\
  Non-thermal Power\tablenotemark{b} & $P_\mr{N}(\geq 25)$ &
  $10^{24}$~erg s$^{-1}$&  $10^{22}-10^{27}$~erg s$^{-1}$\\
 \enddata
 \tablenotetext{a} {Values from part I of this article \citep{mfpart1}; 1 min lower limit
 due to selection effects.}
 \tablenotetext{b} {All parameters are estimated from analysis using 16 seconds
around the time of peak emission in 6-12~keV, so resulting energy
estimates for those times.} \end{deluxetable*}

One further explanation for the unexpectedly large non-thermal power in
the microflares might be that the physical model used is less suited for
small flares than large. The energy in the accelerated electrons was found
using the standard cold thick target model \citep{brown1971}. It has been
suggested that for the lowest energy electrons the target is warm and not
cold, and $5kT$ could be used as an approximate cut-off
\citep{emslie2003}. Unfortunately for the microflares, this gives an even
lower cut-off at around 4--5 keV, which results in a even larger
non-thermal power estimate. Nevertheless the idea of the non-thermal
electron distribution smoothly transitioning into the thermal distribution
seems physically more realistic than a sharp low energy cut-off and may
provide a clue to the physics of the energization of the electron
distribution function. Another change to the model of the hard X-ray
emission would be the inclusion of free-bound emissions
\citep{brown2007}, as our models currently only assume free-free
continuum. Microflares might be the ideal type of flares to study this
seldom-considered mechanism as the resulting additional spectral
features would occur above $10$ keV and would likely have been hidden
by the thermal component in large flares. However the temperatures
required for such free-bound features to be present, $\ge20$ MK
\citep{brown2007}, are not typically observed in {\it RHESSI} microflares.

The analysis presented here is only performed at the time of peak
6-12~keV emission even though the most desirable results would
investigate the full time range of each microflare, producing the total
thermal and non-thermal energy estimates. But as this paper has shown, it
is a considerable undertaking just analyzing this peak time period. The
next steps in this microflare study would be to investigate the biases and
instrumental effects using simulations and making improvements to the
forward-fit model of the emission. Although {\it RHESSI} has  for the first
time allowed the thermal energy and non-thermal power distributions to
be studied systematically in microflares, the uncertainty in the transition
between thermal and non-thermal components highlights the instrumental
effects and biases. We also have suggested the possibility that the
standard hard X-ray model used is wrong. The model and instrumental
effects could be studied by investigating the microflares not only with {\it
RHESSI} but other instruments as well. It may however, require an
instrument with better energy resolution and lower instrumental
background than {\it RHESSI} before we can unambiguously determine the
energy component of microflares and decide whether there is an issue
with the hard X-ray model used.

\section{Acknowledgements} NASA supported this work under grant
NAS5-98033 and NNG05GG16G. I. Hannah would like to thank L. Fletcher
and C. Parnell, as well as the rest of the {\it RHESSI} team, for helpful
discussions and M. Battaglia for providing flare data for comparison.


\end{document}